# RESEARCH

**Open Access**

# Comparative analysis of box-covering algorithms for fractal networks

Péter Tamás Kovács[1], Marcell Nagy[1] and Roland Molontay[1,2*]

*Correspondence:
molontay@math.bme.hu
[2] MTA-BME Stochastics Reseach Group, Budapest, Hungary
Full list of author information is available at the end of the article

**Abstract**

Research on fractal networks is a dynamically growing field of network science. A central issue is to analyze the fractality with the so-called box-covering method. As this problem is known to be NP-hard, a plethora of approximating algorithms have been proposed throughout the years. This study aims to establish a unified framework for comparing approximating box-covering algorithms by collecting, implementing, and evaluating these methods in various aspects including running time and approximation ability. This work might also serve as a reference for both researchers and practitioners, allowing fast selection from a rich collection of box-covering algorithms with a publicly available codebase.

## Introduction

It is a challenging task to develop notions that can capture important features of a real-world network. Popular approaches often focus on the degree distribution of the nodes of a given network, on degree-degree correlations, on shortest path related measures, on the clustering of the network, and on other structural measures (Newman et al. 2011). Various classes of networks, such as small-world networks, scale-free networks, and networks with a strong community structure have attracted a lot of research attention in the past two decades (Molontay and Nagy 2021).

In recent years, there has been an increasing interest in the class of *fractal networks*, since fractality has been verified in several real-world networks (WWW, actor collaboration networks, protein interaction networks) (Song et al. 2005; Wen and Cheong 2021). Moreover, fractality has been associated with many important properties of networks such as robustness, modularity, and information contagion (Rozenfeld et al. 2007). For example, fractal networks are found to be relatively robust against targeted attacks, which may provide an explanation why numerous biological networks evolve towards fractal behavior (Gallos et al. 2007). For a wide range of applications of fractal property in networks, see the work of Wen and Cheong (2021).

Since the notion of fractal networks is motivated by the notion of fractal geometry, the method to identify fractal behavior of networks is similar to that of regular fractal objects: using the *box-covering method*, also called *box-counting method.* The method aims to determine the minimum number of boxes needed to cover the entire network. Although this problem belongs to the family of NP-hard problems (Song et al. 2007), numerous algorithms have been proposed to obtain an approximate solution.

In this contribution, we provide a systematic review and comparative analysis on a large collection of algorithms that have been proposed throughout the years to perform the box-covering of the network. The high number of approximating algorithms introduced recently calls for an impartial, systematic comparison of the algorithms. There are a few recent works that describe the most important box-covering algorithms (Wen and Cheong 2021; Rosenberg 2020; Huang et al. 2019; Deng et al. 2016), however, to the best of our knowledge, this is the first comprehensive comparative study in this field. After collecting and implementing various methods, we compared their performance in terms of approximation ability and running time on a number of real-world networks. In addition, we made all our codes publicly available on GitHub (https://github.com/PeterTKovacs/boxes) together with detailed documentation and a tutorial Python notebook. It makes our results reproducible and allows for future contributions from interested community members.

**Fractality in geometry**

The term fractal was coined by Benoit B. Mandelbrot and the concept was originally developed for sets in a Euclidean space (Mandelbrot 1982; Falconer 2004). In fractal geometry, the box-covering algorithm is one possible way to estimate the fractal dimension of a set $S$ in a $d$-dimensional space ($\mathbb{R}^d$). The set is lying on an evenly spaced $d$-dimensional grid and the hypercubes of this grid are boxes of size $l$. We consider the $N(l)$ minimal number of boxes of size $l$ that are needed to cover the set and see how it scales with the box size. The box-counting or Minkowski dimension is defined as

$$\dim_B(S) = \lim_{l \to 0} \frac{\log N(l)}{-\log l}$$

The exact mathematical definition of fractals goes beyond the scope of this paper. Roughly speaking, a set with a non-integer box-counting dimension is considered to have fractal geometry, since it suggests that the fractal scales differently from the space it resides in.

**Fractality of networks**

The notion of fractal networks is motivated by the concepts from fractal geometry and measuring the fractal dimension of a network is analogous to the geometric case (Song et al. 2005). The box-counting method can be easily generalized to networks since the vertex set of an undirected graph together with the graph distance function (geodesic distance) form a metric space, and the box-counting algorithm generalizes to any metric space. For networks, the *box-covering* also called *box-counting* method works as follows:

- We say that a group of nodes fit into a box of size $l_B$ if, for any pair of these nodes, their shortest distance is less than $l_B$.

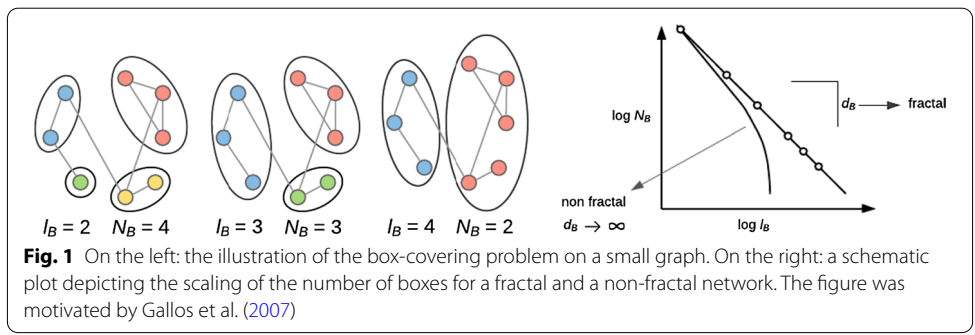

**Fig. 1** On the left: the illustration of the box-covering problem on a small graph. On the right: a schematic plot depicting the scaling of the number of boxes for a fractal and a non-fractal network. The figure was motivated by Gallos et al. (2007)

- A network is covered by boxes of size $l_B$ if its nodes are partitioned such that every group fits into one box of size $l_B$.
- As the number of possible coverings is finite, there is a minimal number of boxes to cover the network with $l_B$ sized boxes. This is what we denote by $N_B(l_B)$.
- If the minimal number of boxes scales as a power of the box size, i.e., $N_B(l_B) \propto l_B^{-d_B}$, the network is informally defined to be *fractal* with a finite *fractal dimension* or *box dimension $d_B$*.

Although the "$\propto$" scaling relation is not well-defined and cannot strictly hold for finite networks, the box-covering-based fractal dimension can be defined mathematically rigorously for infinite networks and graph sequences (Komjáthy et al. 2019). However, in practice, one usually verifies the fractality of real-world networks by approximating the $N_B(l_B)$ values on a feasible range of $l_B$ box size values and then inspects them on a log-log plot to see if the relationship follows a power-law to a good approximation (see Fig. 1). The left side of Fig. 1 shows an optimal box-covering of a small graph with three different box sizes. The right side of the figure illustrates the scaling of the number of boxes $N_B$ and the box size $l_B$ for a fractal and a non-fractal network.

Although fractality is also defined for weighted networks (Wei et al. 2013), in this work we only focus on unweighted and undirected graphs. We also note that besides the box-covering-based fractal dimension, several other network dimension concepts have been proposed throughout the years, some of them are equivalent under some regularity conditions on the network (Wang et al. 2017). For a survey on fractal dimensions of networks, we refer to Rosenberg (2018, 2020).

## Algorithms

The optimal box-covering of some mathematically tractable network models (e.g. $(u, v)$-flower, hierarchical scale-free graph, Song–Havlin–Makse model, Sierpinski network) can be determined rigorously (Rozenfeld et al. 2007; Komjáthy et al. 2019; Niu and Li 2020), however, the box-covering of real networks cannot be studied analytically hence it must be calculated algorithmically. The box-covering of a network is known to be NP-hard (Song et al. 2007). Hence to have an algorithm of practically acceptable time complexity, one must use approximating methods. As it is usual for semi-empirical methods, there are a good number of different proposals for the task, see Table 1. To give some intuitive summary, we can say that most algorithms follow a greedy strategy where the

**Table 1** The box-covering algorithms

| Type | Algorithm | Abbr | Year | References |
|---|---|---|---|---|
| Classic | Random sequential | RS | 2007 | Kim et al. (2007) and Gao et al. (2008) |
| | Greedy coloring | GC | 2007 | Song et al. (2007) |
| | Merge algorithm | MA | 2010 | Locci et al. (2010) |
| Burning | Compact-box burning | CBB | 2007 | Song et al. (2007) |
| | Modified box counting method* | MBC | 2007 | Kitsak et al. (2007) and Yuan et al. (2017) |
| | Max-excluded mass burning | MEMB | 2007 | Song et al. (2007) |
| | Ratio of excluded mass to closeness centrality | REMCC | 2016 | Zheng et al. (2016) |
| | MCWR algorithm | MCWR | 2019 | Liao et al. (2019) |
| Meta-heuristic | Edge-covering with simulated annealing* | ECSA | 2007 | Zhou et al. (2007) |
| | Simulated annealing | SA | 2010 | Locci et al. (2010) |
| | Differential evolution | DE | 2014 | Kuang et al. (2014) |
| | Particle swarm optimization | PSO | 2015 | Kuang et al. (2015) |
| | Multi-objective particle swarm optimization* | MOPSO | 2017 | Wu et al. (2017) |
| | Max–min ant colony optimization algorithm* | ACO | 2017 | Li et al. (2017) |
| Overlapping | Fuzzy box-covering | Fuzzy | 2014 | Zhang et al. (2014) |
| | Overlapping box covering algorithm | OBCA | 2014 | Sun and Zhao (2014) |
| | Improved overlapping box covering* | IOB | 2020 | Zheng et al. (2020) |
| Sampling-based | Minimal-value burning* | MVB | 2012 | Schneider et al. (2012) |
| | Sampling-based method | SM | 2019 | Wei et al. (2019) |
| Weighted | Coulomb's law based box-covering* | CL | 2016 | Zhang et al. (2016) |
| | Deterministic box-covering algorithm* | DBCA | 2020 | Gong et al. (2020) |
| Other | Sketch-based box-covering* | Sketch | 2016 | Akiba et al. (2016) |
| | Community-structure-based method* | COM | 2021 | Giudicianni et al. (2021) |

The starred algorithms are not yet available in our repository distinction between algorithms is drawn by the actual greedy decision method. However, we present a number of approaches beyond the greedy paradigm.

It is also important to mention that there are algorithms that use diameter-based boxes, while other methods use radius-based boxes, also called balls of radius $r_B$ around a center node $c$. The two approaches sometimes cause some ambiguity in the terminology.

Using radius-based boxes require the notion of "centers" or "seeds", which means that each box is assigned to a special, central node. While this idea is natural in $\mathbb{R}^n$, it may be misleading for a graph since the edges between vertices generally cannot be embedded into a Euclidean space. For example, in a box that contains a cyclic subgraph, there is no vertex that would be special by any means. The reason, however, why these nodes are called seeds or center nodes is that these nodes are the first elements of the boxes and the boxes are built around them.

**General remarks**

In the following part, we describe and evaluate the implemented algorithms in detail. Throughout this paper, we denote the box size by $l_B$, the box radius (if applies) by $r_B$, where $l_B = 2r_B + 1$. However, let us note that in the literature the definition of an $l_B$-sized box can be inconsistent. In the *implementation*, we define boxes such that for box-size $\tilde{l}_B$ the distance between two nodes of the same box can be *less than or equal* to $\tilde{l}_B$.

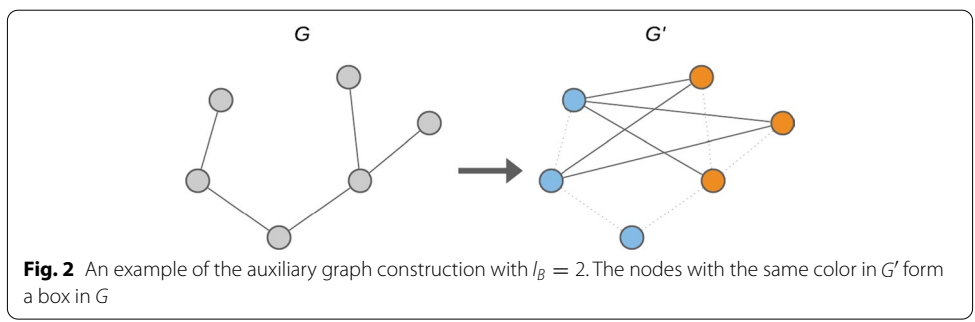

**Fig. 2** An example of the auxiliary graph construction with $l_B = 2$. The nodes with the same color in $G'$ form a box in $G$

In the *evaluation* part, we followed the widely used convention, which means that in an $l_b$-box the distance between any two nodes is *strictly less* than $l_B$. This means that the "equivalent box-size" is $l_B = \tilde{l}_B + 1 = 2r_B + 1$.

In this section, we use the *less than or equal* ("implementation") convention for the algorithms. We will also provide a sketch of the pseudocode for the discussed algorithms. The aim is to foster understanding, there may be minor differences compared to the actual Python code.

As it is apparent from the definition, the box-covering process of a network requires the ability to determine the shortest path between two nodes to be able to decide if they can be in the same box. This could be done in multiple ways, for example, by performing a breadth-first search on the fly. However, for convenience and to avoid calculating the pairwise distances multiple times, we implemented the following approach: in the beginning, we calculate and store the pairwise shortest distances $d(v_i, v_j)$ for all $v_i, v_j \in V$, where $V$ is the vertex set of the graph. This data is then used in the subsequent calculations. With this notion, we define the *ball of radius $r_B$ around center $c$*, which is $B(c, r_B) = \{d(v_i, c) \leq r_B \mid v_i \in V\}$. This term will be used extensively throughout this paper since many algorithms operate on these balls. Note that in this work, the terms graph and network are used interchangeably.

**Greedy coloring (GC)**

The problem of box-covering can be mapped to the famous problem of graph coloring (Song et al. 2007). Therefore, a graph coloring algorithm can be utilized to solve the box-covering problem. The idea is the following: let us consider the *auxiliary graph* of our original network, which consists of the same vertices and "auxiliary edges". This means that two vertices are connected by an edge in the auxiliary graph if and only if their distance in the original network is greater than $l_B$, which means that they cannot be in the same box. The idea is illustrated in Fig. 2, moreover, the pseudocode is detailed in Algorithm 1.

Once we construct the auxiliary graph for a given $l_B$, we perform a graph (vertex) coloring on the auxiliary graph, i.e., we assign colors to the vertices of the auxiliary graph such that (1) adjacent vertices cannot have the same color, (2) and we seek to use the least number of colors. This problem is equivalent to the box-covering of the original network since the assigned colors in the auxiliary graph correspond to the assigned boxes in the original network. The vertices of the same color in the auxiliary graph are at most $l_B$ far away from each other, hence they form a box in the original network. Moreover, the minimum number of colors required to color the vertices (the chromatic number) of the auxiliary graph equals the minimum number of boxes needed to cover the original graph.

---

**Algorithm 1** Auxiliary graph

---

**Require:** $G = (V, E)$, $l_B$,
　$V' = V$ # the nodes of $G'$ are the same
　$E' = []$ # initialize $E'$ as an empty list
　**for** $v_1$ in $V'$ **do**
　　**for** $v_2$ in $V'$ **do**
　　　**if** $d(v_1, v_2) > l_B$ in $G$ **then**
　　　　add $(v_1, v_2)$ edge to $E'$
　　**end for**
　**end for**
　**return** $G' = (V', E')$ auxiliary (simple) graph

---

Unfortunately, the complexity of the graph coloring problem is NP-hard. A well-known approximating algorithm is *greedy coloring*. This algorithm consists of two main steps (Kosowski and Manuszewski 2004):

1. Order the nodes by some method
2. Iterate over the above sequence: assign the smallest possible color ID to every node

The schematic way the greedy coloring box-covering algorithm works is presented in Algorithm 2.

---

**Algorithm 2** Greedy coloring

---

**Require:** $G = (V, E)$, $l_B$, and strategy $s$
　$G'$ = auxiliary_graph($G, l_B$) # see Algorithm 1
　$V^{sorted}$ = order($V, s$)
　**for** $v$ in $V^{sorted}$ **do**
　　assign lowest allowed color ID to $v$ in $G'$
　**end for**
　**return** (node, color) pairs

---

The algorithm is completely well-defined if the strategy, i.e., the ordering method is specified. In this work, we used a random sequence. Even though more advanced methods exist, we think that this *naive* approach serves as a great baseline for the other algorithms.

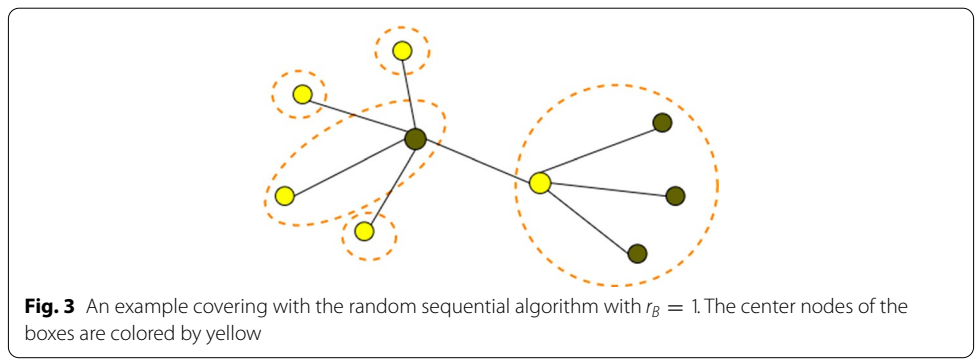

**Fig. 3** An example covering with the random sequential algorithm with $r_B = 1$. The center nodes of the boxes are colored by yellow

Our implementation relies on the greedy coloring implementation of the `networkx` package (Hagberg et al. 2008) and in our Python package, we made it possible to select any other built-in strategy.

Li et al. (2017) introduced a related box-covering algorithm, which uses a so-called max-min ant colony heuristic optimization method to color the vertices of the auxiliary graph.

While in this work we focus on unweighted networks, to be comprehensive, let us note that there are a few box-covering methods, that aim to solve the box-covering problem by assigning weights to the edges of an originally unweighted network. For example, Zhang et al. (2016) proposed a greedy coloring method where the construction of the auxiliary graph is different. They adopted Coulomb's Law to assign *repulsive force*, i.e., weights to the edges, which was motivated by the observation that in real fractal networks hubs are less likely to be connected (Song et al. 2006), however, there are some counterexamples as well (Kuang et al. 2015; Nagy 2018). The construction of the auxiliary graph hence is based on the weighted shortest paths, called smallest repulsive force paths, and not on the original shortest paths. Recently, Gong et al. (2020) introduced a deterministic box-covering algorithm, which is a modification of the algorithm of Zhang et al., namely it starts the coloring with the high-degree nodes.

**Random sequential**

The random sequential algorithm was introduced by Kim et al. (2007) as one of the first approximating box-covering algorithms and similarly to the algorithms that were proposed by Song el al. (2007), it also uses the idea of burning (breadth-first search). Burning means that the boxes are generated by growing them from one randomly selected center (or seed) vertex towards its neighborhood, see Fig. 3. Moreover, once some nodes have been assigned to a box, they are "burned out".

In every step, an unburned center node $c$ is randomly chosen, and then the ball of radius $r_B$ is formed around $c$, more precisely, the algorithm selects those unburned nodes that are at most $r_B$ far from the center node $c$. These newly burned nodes together form a new box. Note that in the original paper, the center node is selected from the whole set of nodes $V$, yet we modified it to select it from the uncovered set of nodes. The authors argue that in some cases, that was necessary to obtain the desired behavior, namely, they stated that if they disallow *disconnected boxes*, they get different results for the scaling of $N_B$ on the graph of WWW. In a disconnected box, the vertices are not connected, but there is a path between them and the center node, which is however not a member of the box because it has been burned already.

---

**Algorithm 3** Random sequential (RS)

**Require:** $G = (V, E)$, $r_B$
   $U = V$    # uncovered nodes
   $boxes = [\ ]$    # list of the created boxes that will be returned
   **while** $U \neq \emptyset$ **do**
      $c = \text{random\_choice}(U)$    # randomly chosen uncovered center node
      append $U \cap B(c, r_B)$ to boxes
      $U = U \setminus B(c, r_B)$
   **end while**
   **return** $boxes$

---

One year later, Gao et al. (2008) proposed a so-called rank-driven box-covering algorithm, which is a computationally more efficient version of the random sequential algorithm. Zhang et al. (2017) proposed a modified version of the random sequential algorithm, where the seeds of the firstly selected boxes are the nodes with the highest degrees, and it forces that the largest hubs are in different boxes. The authors also compare their algorithm to the original random sequential method, and a so-called P-BC variant of the RS, which selects always the highest degree nodes as seeds.

**Compact box burning (CBB)**

As the name suggests, similarly to the random sequential algorithm the compact box burning algorithm also uses the concept of burning. The algorithm was introduced by Song et al. (2007) together with the greedy and the MEMB algorithms. The main point is to grow boxes by picking new nodes randomly from a candidate set *C* that contains all nodes that are not farther away than $l_B$ from any node that is already in the candidate set. In the end, this set *C* is going to form a box, and the process guarantees that the box is compact.[1]

---

**Algorithm 4** Compact Box Burning (CBB)

**Require:** $G = (V, E)$, $l_B$
   $U = V$    # for uncovered nodes
   $boxes = [\ ]$    # list for assigned boxes
   **while** $U \neq \emptyset$ **do**
      $box = [\ ]$    # list for current box
      $C = U$    # candidates for box members
      **while** $C \neq \emptyset$ **do**
         $p = \text{random}(C)$
         append $p$ to $box$
         $C = C \setminus \{d(p, v_i) > l_B \mid v_i \in V\}$    # remove $p$ incompatible nodes from $C$
      **end while**
      append $box$ to $boxes$
      $U = U \setminus box$
   **end while**
   **return** $boxes$

---

[1] Roughly speaking, we cannot add any more nodes.

Kitsak et al. (2007) introduced a simplified version of the CBB algorithm which estimates the number of boxes rather fast and it has been also used and detailed in a more recent work (Yuan et al. 2017). The modified algorithm does not construct the actual boxes, it only approximates $N_B$ for a given $l_B$ as follows: first initializes $N_B$ and sets its value to zero. Then it selects a random center node $c$, marks the nodes of the ball of radius $l_B$ centered on $c$, and finally increases the value of $N_B$ by one. Then randomly selects another unmarked center node and repeats this process until the whole network is covered. The authors stress that this estimation of $N_B$ is always less than the actual minimum number of boxes needed to cover the network, hence to improve the estimation they suggest performing the computation many times and then taking the maximum of the estimations which is claimed to be a better approximation of $N_B$ and that still has a small time complexity.

Note that this algorithm could be also considered as a simplified *sampling-based* box-covering method (see "Sampling-based method" section), which is a more general framework that selects the "best" box-covering from many independent box covers generated by a simple covering algorithm.

### Maximal excluded mass burning (MEMB)

The Maximal Excluded Mass Burning algorithm is the third box-covering algorithm that was presented in the influential paper of Song et al. (2007). Instead of using $l_B$, this method also uses the notion of radius $r_B$ and centered boxes: every box has a special node, a center. Boxes are constructed such that every member node of the box is not farther away than $r_B$ from the center node. The algorithm guarantees that all the boxes are connected, i.e., two nodes of the same box can always be reached with a path that is inside of the box.

The algorithm could be interpreted as an improvement of the random sequential method in some aspects but the analogy fails because the way burning is implemented is different from random sequential. It works as follows:

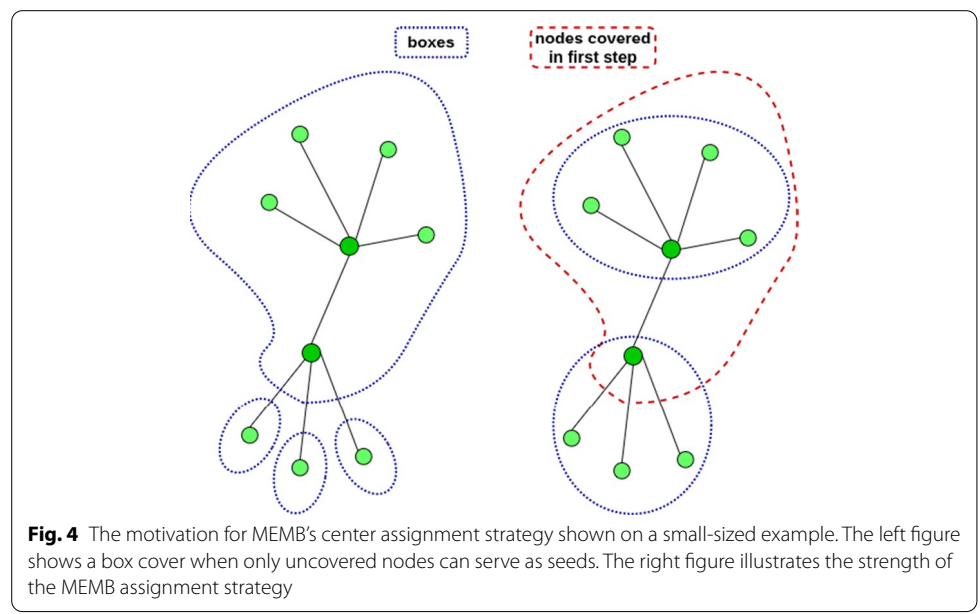

**Fig. 4** The motivation for MEMB's center assignment strategy shown on a small-sized example. The left figure shows a box cover when only uncovered nodes can serve as seeds. The right figure illustrates the strength of the MEMB assignment strategy

- First, the next center is chosen on the basis of *maximal excluded mass*, that is the number of uncovered nodes not farther away than $r_B$ from the center node.[2]
- Once the next center is chosen, all uncovered nodes within $r_B$ are covered, but not yet assigned to boxes.
- Finally, after every node is covered, non-center nodes are assigned to centers in a way that the resulting boxes are connected.

The pseudocode of the MEMB algorithm is presented in Algorithm 4.

---
**Algorithm 5** Maximal Excluded Mass Burning (MEMB)
---
**Require:** $G = (V, E)$, $r_B$
  $U = V$  # uncovered nodes
  $C = \{\}$  # center nodes
  **while** $U \neq \emptyset$ **do**
    calculate $m_i$ excluded mass for non-center nodes
    let $p$ be the node for which $m_p$ is maximal
    $C = C \cup \{p\}$
    $U = U \setminus B(p, r_B)$
  **end while**
  let $c_{dist}$ contain the distance to the closest center for every node
  let $C'$ be the list of non-center nodes sorted by their $c_{dist}$ value
  **for** $n$ *in* $C'$ **do**
    $m =$ find a random neighbor of $n$ whose $c_{dist}$ is smaller
    allocate $n$ to the center node which is the center of $m$
  **end for**
  **return** the formed boxes that are identified by their center nodes

Figure 4 helps to understand the reason why boxes are only formed in the end and not burned on the fly. The authors' motivation was that in scale-free networks, where there are a few hubs, the hubs should be selected as the center of the boxes, otherwise the tiling of the network is going to be far from the optimal case. That is why the algorithm first selects centers and then assigns the remaining nodes to them.

**Ratio of excluded mass to closeness centrality (REMCC)**

The REMCC box-covering algorithm has been introduced by Zheng et al. (2016), and it can be regarded as a modification of the MEMB algorithm. Both algorithms rely on excluded mass, but it also considers the ratio of excluded mass to closeness centrality (REMCC) to select the center nodes.

In every step, a new center is chosen from the uncovered set of nodes such that the chosen node has the maximal *f* score, which is a novel metric in the paper. The *f* score is defined as the ratio of the excluded mass and closeness centrality or in other words, it is the product of the excluded mass and the average length of the shortest paths to all other nodes. The authors argue that the reason why they consider the closeness centrality in the selection of the center nodes is that if we choose a central (important) node of the network as a center (seed) of a box, then eventually more boxes will be needed to cover the network. After the center node is selected, all nodes in the $r_B$ ball of the center are covered. In this procedure—in contrast to the MEMB algorithm—centers are selected from the uncovered set of nodes.

---
[2] Including the center node as well, if it is uncovered yet. It is also possible to have a covered node as the next center.

The steps of the REMCC box-covering algorithm are detailed in Algorithm 6. Note that in the current implementation, the algorithm does not return the formed boxes, only the centers are determined, however, after the center nodes are given, the box-building method of the MEMB algorithm can be applied.

---

**Algorithm 6** Ratio of Excluded Mass to Closeness Centrality (REMCC)

**Require:** $G = (V, E)$, $r_B$
  $U = V$    # uncovered nodes
  $C = \{\}$    # centers
  calculate $l_i$: the average shortest path from $i$ to other nodes
  **while** $U \neq \emptyset$ **do**
    calculate $m_i$ excluded mass for the uncovered nodes
    let $p$ be the node for which $f_p = m_p \cdot l_p$ is maximal
    $C = C \cup \{p\}$
    $U = U \setminus B(p, r_B)$
  **end while**
  **return** $C$

---

### MCWR

Recently, Liao et al. (2019) proposed an algorithm, called MCWR, which is a combination of the MEMB and the random sequential (RS) algorithms. Their goal was to create an algorithm that has the accuracy of the MEMB and the fast speed of the random sequential algorithm. In addition, to reduce the time complexity of the MEMB algorithm, a new way of keeping track of the excluded mass is proposed.

The authors argue that due to the excluded-mass-based selection of the central nodes, the MEMB algorithm has high accuracy but also high time complexity. Since the random sequential algorithm selects the center nodes randomly, it is one of the fastest algorithms but it gives a very rough approximation of the optimal number of boxes.

The main idea of the MCWR algorithm is that it mixes the two ways of center node selection. More specifically, the authors introduced a parameter $p \in [0, 1]$, which represents the mixing portion of the MEMB method in the approach of choosing a center node. Hence, before choosing a new center, a biased coin is tossed and with probability $p$ the algorithm performs the usual MEMB steps and with probability $1 - p$ it chooses a center uniformly at random.[3] After finding a new center, the excluded mass of the vertices is updated, but in contrast to the original MEMB algorithm, to reduce the time complexity, it only updates the mass for the nodes that are inside of the newly covered ball of radius $r_B$. The assignment of the center nodes to the non-center nodes, i.e., the construction of the actual boxes, is the same as in the MEMB algorithm.

Note that the setting of the mixing parameter $p$ is indeterminate. The authors apply the algorithm with different $p$ settings to real networks without a suggested default setting. However, it is clear that if $p = 0$, the MCWR is the same as the RS, and similarly if $p = 1$, it works like the MEMB algorithm. Therefore, the smaller the $p$ value is, the "closer" the algorithm is to the random sequential, hence the faster and less accurate the MCWR is.

---

[3] Here, the choice is made such that covered nodes except the ones with $m_{ex} = 0$ are allowed.

---

**Algorithm 7** MCWR

**Require:** $G = (V, E)$, $r_B$, $p$
    $U = V$     # uncovered nodes
    $C = \{\}$     # centers
    calculate $m_i$ excluded mass for all nodes
    **while** $U \neq \emptyset$ **do**
        **if** $UNI[0, 1) < p$ **then**
            let $v$ be the node for which $m_v$ is maximal
        **else**
            let $v$ be a random non-center node
            **while** $B(v, r_B)$ has only marked nodes **do**
                let $v$ be a random non-center node
            **end while**
        **end if**
        $N = U \cap B(v, r_B)$     # newly covered nodes
        **for** $n$ $in$ $N$ **do**
            **for** $l$ $in$ $B(n, r_B)$ **do**
                decrease $m_l$ (by one)
            **end for**
        **end for**
        $C = C \cup \{v\}$
        $U = U \setminus B(v, r_B)$
    **end while**
    let $c_{dist}$ contain the distance to the closest center for every node
    let $C'$ the list of non-center nodes sorted by their $c_{dist}$ values
    **for** $n$ $in$ $C'$ **do**
        $m = $ find a random neighbour of $n$ whose $c_{dist}$ is smaller
        allocate $n$ to the same center as of $m$'s
    **end for**
    **return** the established boxes identified by centers

---

**Merge algorithm (MA)**

The merge algorithm (MA) has been introduced by Locci et al. (2010), who investigated the fractal dimension of a software network using the merge algorithm, the simulated annealing, and the greedy coloring box covering algorithms.

The merge algorithm is based on the following simple idea: first, every node itself is a box (the authors refer to the boxes as clusters), and then for $l_B > 0$ the boxes are formed using a successive aggregation approach, i.e., two boxes are merged if their distance is at most $l_B$, more precisely, if the size of the union of the boxes is still at most $l_B$. Thus, to perform the box-covering of a network with box size $l_B = k + 1$, the Merge algorithm uses the previous results, i.e., the cover of the network with $l_B = k$ sized boxes. The pseudocode of the merge algorithm is detailed in Algorithm 8. Note that Locci *et al.* refers to the boxes as clusters, that is why in the code they are denoted by *c*.

---

**Algorithm 8** Merge algorithm

**Require:** $G = (V, E)$, $l_B$
$\quad C = [\{v_i\} \mid v_i \in V]$
$\quad l = 1$
$\quad$**while** $l \leq l_B$ **do**
$\quad\quad C' = [\,]$
$\quad\quad$**while** $C \neq \emptyset$ **do**
$\quad\quad\quad c_i = \mathsf{random}(C)$
$\quad\quad\quad$ find all $c_j$ $(i \neq j)$ such that $c_i \cup c_j$ is a valid box of size $l$
$\quad\quad\quad$**if** $c_j$'s exist **then**
$\quad\quad\quad\quad$choose a random $c_j$
$\quad\quad\quad\quad$append $c_i \cup c_j$ to $C'$
$\quad\quad\quad\quad$remove $c_i$, $c_j$ from $C$
$\quad\quad\quad$**else**
$\quad\quad\quad\quad$append $c_i$ to $C'$
$\quad\quad\quad\quad$remove $c_i$ from $C$
$\quad\quad\quad$**end if**
$\quad\quad$**end while**
$\quad\quad C = C'$
$\quad\quad l = l + 1$
$\quad$**end while**
$\quad$**return** $C$

---

**Simulated annealing (SA)**

The Simulated Annealing (SA) box-covering algorithm has also been introduced by Locci et al. (2010). Generally, simulated annealing is a meta-heuristic technique that was inspired by the annealing in metallurgy and it is used in optimization problems for approximating the global optimum in a large (typically discrete) search space, hence it can also be applied to approximate the minimum number of boxes needed to cover a network. In the simulated annealing context, the state of the physical system is the box covering of the network and the "internal energy" of the state is the number of boxes $N_B$.

In the simulated annealing process, several states, i.e., coverings are checked, and a new neighbor state $S'$ of the current state $S$ is created using the following three operations: (1) moving a single node from one box (with at least two nodes) to another if the size of the extended box is still less than $l_B$, (2) creating new box by excluding a node from a box, and (3) merging boxes using the merge algorithm (see Algorithm 8).

After performing these steps in the above sequence, the internal energy $E(S')$ of the new state $S'$, i.e., the new approximation of the number of boxes is checked, and if $E(S') \leq E(S)$ then the algorithm continues with the new configuration. However, if the new boxing is worse than the previous one, i.e., $E(S') > E(S)$ then with probability $p = \exp\left(-\frac{\Delta N_B}{T}\right) = \exp\left(-\frac{E(S')-E(S)}{T}\right)$ it is still accepted. Note that at each iteration, the system is cooled down, which means that the probability $p$ of accepting a worse covering is reduced: $T' = T \cdot c$, where $c < 1$ is the cooling constant parameter.

We introduced several modifications in the implementation compared to the original paper. First, the operations (1) and (2) are rather vaguely defined in the paper. To be able to perform these operations, it must be ensured, for example that after the removal of a given node, the original box would remain nonempty. Thus, the changes compared to the paper of Locci et al. (2010) are as follows:

- *Moving nodes* we only try to move nodes into neighboring boxes to spare iterating over all nodes when checking the distance ($l_B$) condition. Moreover, instead of ensuring that $k_1$ movements are performed, we perform $2 \cdot k_1$ trials, which may fail if for a randomly chosen box we cannot move any additional node from the neighbor boxes. If $k_1$ movements are successfully performed, the process is terminated.
- *Creating new boxes* we proceed in an analogous manner, we perform at most $k_2$ trials on creating new boxes: a random box is chosen and if it has at least two nodes then a random node of this box is removed and that node forms a new box on its own. If $k_2$ creations are successfully performed, the process is terminated.
- *Merging boxes* as opposed to the original paper (Locci et al. 2010), in our implementation this step is performed before evaluating the energy of the new state, i.e., the number of boxes are counted after the merge procedure is also done.

Our implementation of the simulated annealing box-covering algorithm is detailed in Algorithm 9.

---

**Algorithm 9** Simulated Annealing (SA)

**Require:** $G = (V, E)$, $l_B$, $k_1$, $k_2$, $k_3$, $T_0$, $c$
  $T = T_0$
  $S = \mathbf{merge}(G, l_B)$   # the initial boxes are produced by the merge algorithm
  **for** $i = 1$ to $k_3$ **do**
    $S' = S$
    **try** at most $2 \cdot k_1$ times:
      choose a $b$ random box
      try to find a $v_a \notin b$ node that has a neighbor in $b$,
        and that can be moved into $b$ and its box has at least 2 members
      **if** $v_a$ exists, move it into $b$ (in $S'$)
      **if** moved $k_1$ nodes successfully: stop trying
    **try** at most $2 \cdot k_2$ times:
      choose a $b$ random box
      **if** it has at least 2 nodes: remove one random and make it to a new box in $S'$
      **if** moved $k_2$ nodes successfully: stop trying
    merge boxes in $S'$ using the Merge Algorithm 8.
    **if** $E(S') \leq E(S)$ or $UNI[0,1) < \exp\left(-\frac{E(S')-E(S))}{T}\right)$ **then**
      $S = S'$
    **end if**
    $T = c \cdot T$
  **end for**
  **return** $S$

---

Zhou et al. (2007) proposed an edge-covering method, where the covering is performed with the simulated annealing optimization. Clearly, an edge-covering approach automatically covers the nodes as well, but the estimated number of boxes is greater than or equal to the number of boxes returned by a node-covering method since in the edge-covering method some nodes may be covered multiple times. Thus, the edge-cover method also estimates a different box-dimension of the network.

**Overlapping-box-covering algorithm (OBCA)**

The overlapping-box-covering algorithm was introduced by Sun and Zhao (2014). As the name of the algorithm suggests, the novelty of this method is that instead of partitioning the nodes, it uses overlapping boxes. The authors claim that this technique yields a more robust estimation of the number of boxes required to cover the network because separated boxes are prone to randomness. The authors suggest that instead of burning boxes on the fly, one should only mark possible boxes while processing the data and then choose the final boxing in the end.

The algorithm proceeds by first only creating box proposals, during which possible center nodes are iterated in ascending order with respect to the degree.

- A node is a possible center if it is uncovered yet. The algorithm starts with the small-degree nodes and large-degree nodes are checked at last.
- In a "proposed box", those nodes are included whose distance from each other is at most $l_B$. They are chosen from nodes whose distance from the center is at most $l_B$.
- Once a proposed box is formed, the "covered frequency" of all nodes belonging to the newly proposed box is increased by one. The covered frequency of a node denotes how many overlapping proposed boxes contain it.

The final step of the algorithm is the revision of the proposed boxes. A box is called "redundant" if it only contains nodes that are also contained by other boxes, i.e., the covered frequency of all the nodes in a redundant box is larger than 1. In the last step, the redundant boxes are deleted and the covered frequency of the nodes of the recently removed box is decreased by 1. After the iteration is over, only non-redundant boxes remain. The detailed pseudo-code of the overlapping-box-covering method is described in Algorithm 10.

---

**Algorithm 10** Overlapping Box Covering Algorithm (OBCA)

**Require:** $G = (V, E)$, $l_B$
  $boxes = [\,]$    # list (ordered sequence)
  let $f$ contain the covering frequency for each node, initialized with $0$
  let $N$ contain the nodes sorted ascending by degree
  **for** $n$ in $N$ **do**
    **if** $f[n] > 0$ **continue**
    $box = B(n, l_B)$
    let $C$ be $box$, sorted ascending by $f[\cdot]$
    **for** $i = 1$ to $|C|$ **do**
      **if** $C[i] \notin box$ **continue**
      **for** $j = i + 1$ to $|C|$ **do**
        **if** $d(C[i], C[j]) > l_B$ **then**
          **if** $C[j] = n$ **then**
            remove $C[i]$ from $box$
            **break**
          **else**
            remove $C[j]$ from $box$
          **end if**
        **end if**
      **end for**
    **end for**
    **for** $b$ in $box$ **do**
      $f[b] +\!= 1$
    **end for**
    append $box$ to $boxes$
  **end for**
  **for** $box$ in $boxes$ **do**
    **if** all $b \in box$ have $f[b] > 1$ **then**
      remove $box$ from $boxes$
      $f[b] -\!= 1$ for all $b \in box$
    **end if**
  **end for**
  **return** $boxes$

---

In a recent work, Zheng et al. (2020) proposed a modification of this algorithm, which they refer to as the improved overlapping box covering algorithm (IOB). The authors argue that the algorithm could achieve better results if it would rank and sort the nodes according to the excluded mass of closeness centrality instead of the degree.

**Differential evolution (DE)**

The differential evolution (DE) box-covering algorithm has been introduced by Kuang et al. (2014). In general, differential evolution is also a metaheuristic optimization method that belongs to the class of evolutionary algorithms, and it does not require the optimization problem to be differentiable since it optimizes a problem by iteratively trying to improve a candidate solution.

The proposed differential evolution box-covering algorithm uses the deterministic sequential greedy coloring algorithm such that the coloring sequence of nodes is represented by an *N*-dimensional vector, where *N* is the size of the network. The vector encoding of the different sequential greedy coloring procedures transforms the optimization problem into an *N*-dimensional space, which makes it possible to use the differential evolution paradigm.

Similarly to the simulated annealing, the DE algorithm also performs some operations on the population of these vectors to generate new possible solutions. Namely, there are two operations: mutation and crossover. In the mutation process, new vectors are created using randomly chosen three existing vectors. Moreover, to further increase the diversity of the set of examined solutions, the crossover operation generates new random vectors by randomly mixing the coordinates of the existing vectors and the newcomer vectors that were created in the mutation process. At the end of each iteration, the best solution is saved. The algorithm is detailed in Algorithm 11.

---

**Algorithm 11** Differential Evolution (DE)

**Require:** $G = (V, E)$, $l_B$, $p$, $f$, $c$, $g$
  $G' = \text{auxiliary\_graph}(G, l_B)$
  $X_0 = [[UNI[0,1)^N] \times p]$    # population of $p$ random vectors
  **for** $i = 1$ to $g$ **do**
    # The mutation phase
    **for** $j = 1$ to $p$ **do**
      draw $r_1$, $r_2$, $r_3$ mutually different indices that also differ from $j$
      $v_j = X_{i-1}[r_1] + f \cdot (X_{i-1}[r_2] - X_{i-1}[r_3])$
    **end for**
    # The crossover phase
    **for** $j = 1$ to $p$ **do**
      draw $k$ coordinates from $\{1, 2, ..., N\}$ randomly
      **for** $m = 1$ to $N$ **do**
        **if** $UNI[0,1) < c$ or $m = k$ **then**
          $u_j[m] = v_j[m]$
        **else**
          $u_j[m] = X_{i-1}[j][m]$
        **end if**
      **end for**
    **end for**
    # The selection phase
    **for** $j = 1$ to $p$ **do**
      Let $s_1$ and $s_2$ be the sequence of nodes defined by $u_j$ and $X_{i-1}[j]$ respectively.
      $n_1 = \textbf{greedy}(G, s_1)$    # number of boxes
      $n_2 = \textbf{greedy}(G, s_2)$
      **if** $n_1 < n_2$ **then**
        $X_i[j] = u_j$
      **else**
        $X_i[j] = X_{i-1}[j]$
      **end if**
      $best = $ save the best boxing from $s_1$ and $s_2$ if it is better than the current global best
    **end for**
  **end for**
  **return** $best$

**Particle swarm optimization (PSO)**

A box-covering method using a heuristic optimization method called discrete Particle Swarm Optimization was presented by Kuang et al. (2015). The authors argue that the differential evolution algorithm (DE), introduced in their previous work (Kuang et al. 2014), has two main drawbacks that can be improved by the PSO algorithm. First, the DE algorithm has a continuous search space which increases the computational complexity. Moreover, it extensively uses the greedy algorithm which increases the time complexity.

To be able to use the PSO optimization method, the box-covering problem has to be encoded in a discrete form using the *position* and *velocity* of so-called "particles". A particle represents a valid boxing of the network, more precisely, if the size of the network is $N$, then the position of the particle is an $N$-dimensional vector $X = (x_1, \ldots, x_N)$ where the $i$th coordinate $x_i$ denotes the ID of the box where the $i$th node belongs to, moreover $x_i \in 1, 2, \ldots, N$. The velocity of a particle is an $N$-dimensional binary vector $V = (v_1, \ldots, v_N)$ where $v_i$ indicates whether the $i$th node is ready to change its box, i.e., the value of $x_i$ can be changed.

The algorithm operates with a set (swarm) of particles and in every step, each vector (particle) may be updated depending on its velocity, its most optimal boxing, the whole swarm's best boxing, and on random variables as well. In the end, the best overall boxing is returned. The PSO algorithm is detailed in Algorithm 12. Note that we defined some auxiliary functions and operators before the actual algorithm. Unfortunately, the $\oplus$ operator is vaguely defined in the original paper, but we believe that the one below is the most meaningful interpretation of it.

In a follow-up paper, the authors propose a modification of the PSO algorithm (Wu et al. 2017), called multiobjective discrete particle swarm optimization (MOPSO) box-covering algorithm, which aims to solve two optimization problems simultaneously. Besides minimizing the number of boxes required to cover the network, the algorithm maximizes the fractal modularity (Gallos et al. 2007) defined by the boxing, which was motivated by the observation that fractal networks have a highly modular structure (Song et al. 2006). The pseudocode of the MOPSO box-covering algorithm is almost the same as the PSO. The only difference is that in the last part of the code, where it checks whether the current solution is better than the population and global best, it is not enough to check the number of used boxes but a better solution is also required to have a higher modularity (Wu et al. 2017).

---

**Algorithm 12** Particle Swarm Optimization (PSO)

```
# Preliminary definitions

⊕ operator
    x ⊕ y = integer(x ≠ y)     # for integers

def sig(z)     # for numbers
return  1 if UNI[0, 1) < 1/(1+e^-z), and 0 otherwise

def nbest(n, x_i, B)     # node, current box, neighbor boxes
while B ≠ ∅ do
    let x_k be randomly drawn from B
    discard x_k from B
    if x_k = x_i continue
    if d(n, m) ≤ l_B for all m ∈ box(x_k) then
        return  x_k
    end if
end while
return  x_i

# The PSO algorithm
Require: G = (V, E), l_B, g, p, c_1, c_2
P = [encode(greedy(G, l_b, random_sequence_i)) | i = {1, 2, ..., p}]
V = [zeros of shape of P]
P^best = P
let g^best be the best boxing from P
for i = 1 to g do
    for j = 1 to p do
        for k = 1 to |V| do
            # for the sake of simplicity, we write V_jk and P_jk instead of V[j][k] and P[j][k]
            V_jk = sig(UNI[0, 1) · V_jk + UNI[0, 1) · c_1(P^best_jk ⊕ P_jk) + UNI[0, 1) · c_2(g^best[k] ⊕ P_jk))
            if V_jk = 1 then
                P_jk = nbest(k, P_jk, {boxes of neighbors of k})
                if changed box: add the node to the new box, and delete it from the old one.
            end if
        end for
        if number of boxes less than for P^best[j] then
            P^best[j] = P[j]
            if P[j] is the overall best: g^best = P[j]
        end if
    end for
end for
return  g^best
```

---

### Fuzzy algorithm

The fuzzy box-covering algorithm and the corresponding concept of fuzzy fractal dimension were introduced by Zhang et al. (2014). The authors propose a novel scheme for estimating the fractal dimension of a network. Instead of assigning boxes, they introduce a measure to estimate the fraction of the network one box covers on average. For each node, a box of radius $r_B$ around a central node is constructed and the contributions of the nodes of the box are summed. This contribution is a so-called "membership function" that exponentially decays with the distance from the central node. After aggregating and normalizing these contributions, we get an estimation of the proportion of the network that a box covers on average. Taking its inverse gives the approximating box number.

It must be noted, however, that in our experience this approximating number of boxes is not meaningful, only the regressed $d_B$ fractal dimension may be of practical interest. Pseudo-code of the fuzzy box-covering method is presented in Algorithm 13. Note that our pseudo-code is slightly different from the code presented in Zhang et al. (2014) because we assumed that the networks are undirected, hence it is enough to calculate the distance between $v_j$ and $v_k$ only once.

---

**Algorithm 13** Fuzzy algorithm

**Require:** $G = (V, E), \ R = \{r_i\}$
   **for** $r_i$ in $R$ **do**
      $N^{-1}(r_i) = 0$
      **for** $j = 1$ to $|V|$ **do**
         **for** $k = j + 1$ to $|V|$ **do**
            **if** $d(v_j, v_k) \leq r_i$ **then**
$$N^{-1}(r_i) \mathrel{+}= 2 \cdot \exp\left(-\frac{d^2(v_j, v_k)}{r_i^2}\right)$$
            **end if**
         **end for**
      **end for**
      $N^{-1}(r_i) = \frac{N^{-1}(r_i)}{|V|(|V|-1)}$
   **end for**
   **return** $\{N^{-1}(r_i)\}$

---

**Sampling-based method**

Recently, Wei et al. (2019) proposed a novel way to tile a network, that could be considered as an "advanced version" and extension of the overlapping-box-covering algorithm, however, the authors did not compare it to the OBCA algorithm. Note that this proposal of Wei et al. is rather a framework than a particular algorithm.

There are two main stages in the covering process: the first step is the generation of many box proposals, for example by running the random sequential or CBB algorithms—$n$ times. In the second phase, the authors suggest picking some of the proposed boxes in a greedy manner to tile the network. Obviously, a particular realization of the covering is defined by the employed algorithm, since the final cover consists of the boxes that the greedy strategy selected from the box proposals.

Besides CBB and random sequential, the authors proposed a so-called *maximal box sampling* strategy for sampling box proposals which is described in Algorithm 14. Wei et al. also proposed two greedy strategies: *big-box first* and *small-box removal*. These procedures are detailed in Algorithm 15. The authors show that the small-box-removal sampling considerably outperforms the big-box-first strategy and classical algorithms such as MEMB and CBB.

---

**Algorithm 14** The maximal box sampling method

**Require:** $G = (V, E)$, $l_B$, $n$
   $U = V$
   $P = [\,]$
   **for** $i = 1$ to $n$ **do**
      **if** $U \neq \emptyset$ **then**
         let $s \in U$ random
      **else**
         let $s \in V$ random
      **end if**
      $C = B(s, l_B) \setminus \{s\}$
      $box = \{s\}$
      **while** $C \neq \emptyset$ **do**
         let $t \in C$ random
         add $t$ to $box$
         remove $t$ from $C$
         $C = C \cap B(t, l_b)$
      **end while**
      append $box$ to $P$
      $U = U \setminus box$
   **end for**

---

**Algorithm 15** Greedy selection strategies

   # Big-box first strategy

**Require:** $P = [p_i]$ box proposals, $S = [s_i = |p_i|]$ box sizes
   sort $P$ ascending by corresponding $S$ values
   $U = V$
   $R = [\,]$
   **while** $s_l > 0$ with $s_l$ being the size of the last element of $P$ **do**
      append $p_l \cap U$ to $R$
      cover the (uncovered) nodes in $p_l$: $U = U \setminus p_l$
      modify $S$ accordingly: $s_i = |p_i \cap U|$
      sort $P$ ascending by corresponding $S$ values
   **end while**
   **return** $R$

   # Small-box first strategy

**Require:** $P = [p_i]$ box proposals, $S = [s_i = |p_i|]$ box sizes
   # sizes are directly associated to the proposals and do not form an ordered sequence
   sort $P$ ascending by corresponding $S$ values
   $U = V$
   $R = [\,]$
   let $C(n)$ contain the number of box proposals that contain $n \in V$
   **for** $j = 1$ to $|P|$ **do**
      **if** there is $n \in p_j$: $C(n) = 1$ **then**
         append $p_j \cap U$ to $R$
         # cover the (uncovered) nodes in $p_j$:
         $C(n) = 0$ for all $n \in p_j \cap U$
         $U = U \setminus p_j$
         modify $S$ accordingly: $s_i = |p_i \cap U|$
         sort $P[j + 1 :]$ ascending by $S$
      **else**
         $C(n) \mathrel{-}= 1$ for all $n \in p_j$
      **end if**
   **end for**
   **return** $R$

### Further methods

Note that there are a few algorithms that are not implemented in our Python package yet, but for the sake of completeness, we briefly introduce these methods in this section.

Schneider et al. (2012) introduced a box-covering algorithm that is usually referred to as minimal-value burning (MVB) algorithm in the related works. Their algorithm first creates a box around every node, and then it removes the unnecessary boxes. Although their algorithm is said to be nearly optimal, it comes at a price: its computational complexity is high, i.e., for regular networks it scales exponentially with the number of nodes.

Akiba et al. (2016) introduced a sketch-based box-covering algorithm, which uses bottom-$k$ min-hash sketch representation of the boxes and it is based on the reduction of the box-covering problem to a set-covering problem. The authors compare their algorithm to the GC (Song et al. 2007), CBB (Song et al. 2007), MEMB (Song et al. 2007), and the MVB algorithm (Schneider et al. 2012). Akiba et al. also made their implementations of these algorithms publicly available, which can be found at http://git.io/fractality.

Recently, Giudicianni et al. (2021) proposed a community-structure-based algorithm that can be used for networks with uniform degree distribution, i.e., where the nodes have roughly the same amount of connections such as in infrastructure networks (e.g., road, electrical grid, and water distribution networks).

Note that in this work we do not touch upon the concept of multifractality, but for the sake of completeness we mention that the sandbox algorithm, introduced by Tél et al. (1989), can be used to estimate the multifractal dimension of networks (Liu et al. 2015).

### Network data

We have tested the implemented algorithms on numerous real-world networks, many of which are already known from the related literature (see e.g., Song et al. 2005, Wu et al. 2017, Deng et al. 2016). The analyzed networks are listed in Table 2.

Some of the considered networks are directed or contain more than one connected component. Since box-covering algorithms are developed for undirected and connected graphs, our policy was to connect nodes with an undirected edge if there was a link between them in any direction, moreover, we worked with the largest connected component of unconnected networks. We also remark that there were networks containing loops but we believe that they do not have any effect since any node has 0 distance from itself.

We summarize some simple structural features of these networks in Table 3. Besides the usual basic metrics, the value calling for more explanation is the GINI-score. It was introduced by the Italian statistician and sociologist Corrado Gini to quantify wealth inequality in society. This motivated the use of the metric since we wanted to account for the 'inequality of degree distribution' among the vertices of the network. In our work, the GINI-score is defined as follows:

$$\frac{N+1}{N} - 2 \cdot \frac{\sum_{i=0}^{N-1}(N-i) \cdot k_i}{N \cdot \sum_{i=0}^{N-1} k_i}, \tag{1}$$

where $k_i$ stands for the degree of the $i$th node, $N$ is the number of nodes and the sum in the numerator runs over the nodes sorted by their degree in ascending order.

Note that formula (1) gives 0 for "total equality" (all vertices have the same degree) and bounded by $1 + 1/N$, that is practically 1 for large networks.

We note here that the clustering coefficient was calculated with the built-in `networkx.average_clustering` method with default arguments. The modularity was calculated from a partition with the Girvan-Newman method (see `girvan_newman` in `networkx` with default arguments). To calculate the modularity, we used `quality.modularity` from `networkx` with `weight=None`.

## Evaluation

In this section, we turn to the evaluation and comparison of the implemented algorithms on the networks. Before starting the investigations, the reader is reminded that this is the point where we return to the *evaluation* convention for box sizes, see the introduction of "Algorithms" section. The difference between the two conventions might seem very minor and some authors argue that the particular definition has no effect on the fractal scaling, however, Rosenberg (2020) emphasizes that it can indeed make a large difference. Thus, it makes it essential for the evaluation part—especially for the box dimension approximation—to return to the more frequently used convention on the box size.

### Preliminaries

Here we describe the boxing process we applied in our experiments. It is presented in a pseudocode format (see Algorithm 16), however, this pseudocode does not necessarily

---

**Algorithm 16** Evaluation framework

**Require:** $G = (V, E)$, $m$
  Consider $G$'s maximal connected component
  Establish $r_B$ and $l_B$ values for which we box (around 15 different values up to the diameter for $r_B$, for $l_B$ same equivalent sizes)
  Measure and store the time needed for preprocessing (e.g., computing shortest path data, and auxiliary graph construction)
  **for** $alg$ in algorithms **do**
    set random seed
    **for** $b$ in box sizes that apply to $alg$ **do**
      **for** $i = 1$ to $m$ **do**
        Perform a boxing with $alg$ on $G$ with box size $b$ and measure its running time $t$.
        Add the additional time to $t$ that was necessary in the preprocessing state
        Store the obtained number of boxes $N_B$ and the execution time together
      **end for**
      **if** all the $m$ iterations yielded 1 box **then**
        break
      **end if**
    **end for**
    Write the results into log files
  **end for**

---

follow the Python code strictly in all cases.

We note that for the merge algorithm, the actual process of measuring the running time is a bit more complicated since it computes $N_B$ values for all box sizes up to a maximal value (see Algorithm 8). We recorded the execution time plus the preprocessing time for all box sizes, including $l_B = 1$. Furthermore, to avoid counting the preprocessing time multiple times, we retained the total execution time by subtracting the preprocessing time (that is the running time of the algorithm with $l_B = 1$) from the execution time of the $l_B \geq 2$ coverings, and finally, we summed up these "corrected" runtimes.

In the current analysis, we performed the box covering $n = 15$ times for each box size, for each network, for each algorithm. The algorithms and their parameter settings—along with their respective abbreviations—are listed in Table 4. We intended to choose the shorthand notations to be as intuitive as possible.

**Comparison of the number of boxes**

The most obvious way to analyze our results is to prepare $N_B(l_B)$ plots and inspect them. In these plots, we show the mean $N_B$ value (denoted by $\overline{N}_B$) against the box size $l_B$ for each network. We remind the reader that here, box sizes are to be understood in the evaluation convention, as detailed previously, i.e., it doesn't matter if the algorithm is radius-based or diameter-based, the compared boxes have the same equivalent size. To retain comprehensibility, we only plot around 5 algorithms in the same figure, where the algorithms are denoted with different colors and markers. For comparison, we included the greedy algorithm in each plot. Moreover, note that we only plotted for box sizes appearing in the greedy algorithm's output. Figure 5 illustrates the $N_B(l_B)$ plot for the Tokyo metro network with five algorithms, similar plots for all the other networks and algorithms are available in the supplementary material.[4]

Similarly to the related works (Wu et al. 2017; Schneider et al. 2012), we also plot the difference of the $N_B$ values and the number of boxes returned by a baseline algorithm, which is denoted by $N_B^{\text{base}}$. In our work, the baseline $N_B^{\text{base}}$ was chosen to be the minimal (best) value returned by the greedy algorithm (remember that we ran it 15 times for each box size). With this, we plot the difference of the mean $N_B$ and the baseline $N_B^{\text{base}}$, which is denoted by $\Delta \overline{N}_B$. Figure 5 shows the $\Delta \overline{N}_B(l_B)$ plot for the Tokyo metro network with five different algorithms, similar plots for all the other networks and algorithms are available in the supplementary material.

**Comparison with performance scores**

In the following step, we assess the performance of the algorithms with a performance score, defined as the normalized deviation from the baseline—that is the best output by the greedy algorithm—formally:

$$P(l_B) = \frac{N_B(l_B) - N_B^{\text{base}}(l_B)}{N_B^{\text{base}}(l_B)} \qquad (2)$$

Behold that the lower this score is, the better the coverage of the algorithm is. Also note that the score is defined for every boxing output, meaning that we have $n = 15$ scores for each algorithm, box size, and network. The reason behind the definition of this *P* performance score is that these values—expressing relative performance—can be aggregated over box sizes since they are "dimensionless". This is a very desirable property since we want to draw some conclusions from an intimidating amount of data. There is an additional caveat though, we shall define an appropriate interval of box sizes on which we investigate performance scores. The motivation behind limiting the analysis on certain box sizes is that the performance score is not so informative in those cases when the

---

[4] https://github.com/PeterTKovacs/boxes.

**Table 2** The analyzed networks

| Code | Full name, description | References |
| --- | --- | --- |
| phd | CSphd: P.h.D.'s in computer science | Rossi and Ahmed (2015) |
| cel | *C. elegans*: metabolic reactions between substrates | Rossi and Ahmed (2015) |
| soc | Caltech36: social network from Facebook | Rossi and Ahmed (2015) |
| ful | *A. Fulgidus* whole network: cellular network | Wolfram Data Repository (2019a) |
| pol | Polbooks: books about US politics, compiled by V. Krebs | Rossi and Ahmed (2015) |
| dol | dolphins: dolphins social network | Rossi and Ahmed (2015) |
| mou | Mouse brain data, edges represent fiber tracts | Rossi and Ahmed (2015) |
| min | Minnesota: minnesota road network | Rossi and Ahmed (2015) |
| eco | *E. Coli* whole network: cellular network | Wolfram Data Repository (2019b) |
| tok | Tokyo metro: Tokyo underground network | Chen (2017) |

**Table 3** Structural properties of the analyzed networks

| Network | N | E | D | gini | C | Q |
| --- | --- | --- | --- | --- | --- | --- |
| phd | 1025 | 1043 | 28 | 0.43 | 0.00 | 0.35 |
| cel | 453 | 2025 | 7 | 0.49 | 0.65 | 0.03 |
| soc | 762 | 16,651 | 6 | 0.45 | 0.41 | 0.00 |
| ful | 1557 | 3571 | 14 | 0.39 | 0.00 | 0.01 |
| pol | 105 | 441 | 7 | 0.33 | 0.49 | 0.44 |
| dol | 62 | 159 | 8 | 0.33 | 0.26 | 0.38 |
| mou | 213 | 16,242 | 2 | 0.10 | 0.76 | $-0.00$ |
| min | 2640 | 3302 | 99 | 0.15 | 0.02 | 0.38 |
| eco | 2859 | 6890 | 18 | 0.41 | 0.00 | 0.00 |
| tok | 248 | 319 | 36 | 0.22 | 0.02 | 0.45 |

Columns contain: N: number of nodes, E: number of edges, D: network diameter, gini: GINI coefficient of the degree distribution, C: clustering coefficient, Q: modularity whole network is covered with very few boxes. In practice, the tail distribution of the $N_B$ is usually noisy, thus it is pruned from the fitting when the fractal dimension is estimated. Our criterion to consider a box size was that the baseline shall have at least ten boxes. Otherwise even if the difference between $N_B$ and $N_B^{\text{base}}$ is only one, the relative change compared to the value of $N_B^{\text{base}}$ would be misleadingly large.

Following this line of thought, we also normalized the runtimes to assess the average speed of our algorithms. This is done by dividing the runtime of an algorithm by the time that was needed to calculate the $N_B^{\text{base}}$ baseline number of boxes (coming from the output of the best greedy run). By doing this, we can get an overall speed measure for the algorithms.

At last, we are in the position to concisely define the way we compare the algorithms: we plot the mean performance scores versus the mean normalized runtimes with averaging over the accepted $l_B$ values (defined earlier). With this, we get one plot per network where the algorithms are represented by one marker as Fig. 6 shows. To increase information content, the area of the markers is proportional to the empirical variance of the performance score (so the radius of the dots is proportional to the standard deviation). Note that the sizes of markers on the plots for different networks are not to be compared due to varying scaling. Due to the great differences in runtimes, we divided the plots into two sections, with the left having linear and the right having logarithmic horizontal axis (the plots join at 1.1)—see Fig. 6.

The abbreviated name of the algorithms is written next to their corresponding marker. As a general rule, if the marker is barely visible due to the small standard deviation, the labels give guidance about the markers' location. There are some cases, where this would lead to ambiguity, which is avoided by arrows pointing to the position of the marker. Due to the extremely long running time, we had to terminate the simulated annealing and PSO algorithms on the *E. coli* network. Hence, these data points are missing from the set but we can confidently say that their absence does not alter our conclusions. The results on the Minnesota road network and dolphin social network are depicted in Fig. 6, similar figures for all the remaining networks are available in the supplementary material.[5]

Finally, we conclude this section by comparing performance on multiple networks on the same plot. We chose the following real networks: Minnesota road network (min), Tokyo metro (tok), computer science Ph.D.'s collaboration network (phd), *A. fulgidus* network (ful), and *E. coli* cellular network (eco). The reason why we chose these networks is that the number of accepted box sizes was more than one. To make the figures clean, we formed three groups of the algorithms and plotted these groups in separate graphs, the results are shown in Fig. 7. For comparison, the results of the greedy algorithm are always shown. Although we stress that this investigation is insufficient for deciding which algorithms are "the best", we can observe some patterns that will allow us to propose general conclusions in "Discussion and conclusion" section.

**Remark about variances**

In the analysis so far, we used the variance of the $P$ performance score on the acceptance region, which is determined by the number of boxes in the corresponding baseline value. One shall notice that in general, this deviation can come from two sources. First, the *intrinsic* variance of the algorithm, which is the uncertainty of the outcome for a fixed box size (on a given network). The second possible source is that in the acceptance range the average performance of the algorithm varies with respect to the baseline value.

In the plots so far, we accounted for both types of variances, since we considered the variance of all the $P$ scores in the acceptance region. However, it would be also informative to assess the intrinsic variance too, since our baseline is just another approximating algorithm, even if it is an established one. This comparison is done in Table 5 where the mean intrinsic standard deviations and the total standard deviation of the $P$ performance scores are reported (considering scores only for the accepted box sizes). It is interesting to see that the "volatile" algorithms depicted in the third subplot of Fig. 7 behave differently through this lens: for example, the intrinsic deviation is often much smaller than the total deviation for the merge algorithm and especially for the REMCC, however, in the case of the random sequential, they are much closer to each another. Hence the readers should behold the difference between the notions of intrinsic and total deviations and use the appropriate one for their purpose.

---

[5] https://github.com/PeterTKovacs/boxes.

**Approximating the box dimension**

As a final step in the evaluation, we aim to estimate the fractal dimension of the investigated networks. Here we returned to the mean $N_B$ values for each algorithm, and we did not impose the acceptance criterion for the $l_B$ box sizes for this analysis.

We plotted the logarithm of the average $N_B$ versus the logarithm for $l_B$ and tried to fit a linear function to it. This seemingly easy task involved a lot of subjective judgments on whether the relationship could be called linear and on what range of box sizes shall we perform the regression and if there are enough points on the plot to be able to make a confident statement. Behold that fitting linear functions by consulting the plot and adjusting the setup if necessary is a prerequisite for having meaningful results. In practice, the $\log(N_B) - \log(l_B)$ plot is mostly not linear on the whole range. Generally speaking, the characterization of power laws in empirical data is cumbersome, since the tail of the distribution is usually unreliable due to large fluctuations, furthermore, the identification of the range, where the power-law relation holds is difficult (Clauset et al. 2009).

In this experiment, there may be three outcomes: 1) the fit is performed and the relationship is found to be linear, 2) it is realized that the plot is not "linear enough", or 3) the fit was not possible to carry out (missing or too few points: mouse brain (mou), and *E. coli* (eco) networks). In addition to the estimated fractal dimension, we also accounted for the error of the fit (sum of squared errors) with the following expression:

$$SSE = \sqrt{\frac{\sum(y-\hat{y})^2}{(n-2)\cdot\sum(x-\bar{x})^2}}, \quad (3)$$

where $x$ and $y$ stand for the $\log(l_B)$ and $\log(N_B)$ values, respectively. The numerator is the sum of squared residuals of the fit in the denominator, $\bar{x}$ is the mean of $x$ values, $n$ is the number of data points in the fit. Due to the subjective nature of the experiment, the first two authors have performed this fitting procedure independently and then compared the results. Numerical values are reported in Tables 6 and 7.

**Discussion and conclusion**

In this final section, we summarize our observations and point out some directions our analysis could be extended.

**Comparison of algorithms**

We compared the algorithms primarily based on their *P* performance scores that measure relative performance to the baseline values (the best result of the greedy coloring algorithm). The main tool of comparison is to plot the mean *P* scores against the average relative runtime with respect to the baseline for networks with sufficiently large diameters. The radius of the markers is proportional to the standard deviations of the *P* performance score. Firstly, the *P* scores and runtimes shall be considered and the standard deviations should be considered only as a secondary feature due to the multiple sources of variance.

**Table 4** The analyzed algorithms with their respective parameter values

| Abbr. | Algorithm | Hyperparameters |
| --- | --- | --- |
| cbb | cbb | – |
| d30 | Differential evolution | p: 40, f: 0.9, c: 0.85, g: 30 |
| d70 | Differential evolution | p: 40, f: 0.9, c: 0.85, g: 70 |
| fuz | Fuzzy | – |
| gre | Greedy | Strategy: random_sequential (see `networkx` package) |
| mc.25 | mcwr | p: 0.25 |
| mc.5 | mcwr | p: 0.5 |
| mc.75 | mcwr | p: 0.75 |
| memb | memb | – |
| mer | Merge | – |
| obca | obca | – |
| ps.2k | pso | g: 200, p: 99, $c_1$: 1.494, $c_2$: 1.494 |
| ps1k | pso | g: 1000, p: 99, $c_1$: 1.494, $c_2$: 1.494 |
| remcc | remcc | – |
| rs | Random sequential | – |
| sa | Simulated annealing | $k_1$: 5000, $k_2$: 5, $k_3$: 20, $T_0$: 0.6, c: 0.995 |
| sm10 | Sampling | Inner algorithm: maximal box sampling, n: 10, strategy: small first |
| sm40 | Sampling | Inner algorithm: maximal box sampling, n: 40, strategy: small first |
| sr10 | Sampling | Inner algorithm: random sequential, n: 10, strategy: small first |
| sr40 | Sampling | Inner algorithm: random sequential, n: 40, strategy: small first |

The leftmost column stands for the abbreviation. In some plots, the expanded version of these abbreviations appears but it is clear what they mean

The results shown in Figs. 6 and 7 might allow us drawing the following conclusions. First, it should be noted that there are many algorithms that turned out to be way too slow on the analyzed networks: differential evolution (DE), particle swarm optimization (PSO), sampling with maximal box sampling, and simulated annealing (SA). We must also note that these methods yield relatively low box numbers (which is favorable) but the price for this seems to be too heavy. It is also interesting to note that three of these methods (DE, PSO, SA) are metaheuristic algorithms with several hyperparameters. We followed the instructions of the original papers to set the parameters but it seems that tuning the hyperparameters has a high effect on the results and it might be troublesome.

Our results are congruent with that of the related works, for example, Locci et al. (2010) reported the running time of the greedy coloring, merge, and simulated annealing algorithms, and their results also show that the order of magnitude of the runtime of the SA algorithm (1177 s) is twice as much as that of the GC (13 s) and the merge algorithm (21 s) on the *E. coli* network. Furthermore, Akiba et al. (2016) reported that the MEMB algorithm is faster than the greedy coloring and the CBB algorithms, however, in our experiments the difference between the time complexity of the CBB and MEMB algorithm was smaller.

Secondly, there is a group of relatively inaccurate methods as random sequential, REMCC, merge, and sampling with random sequential. Although these algorithms have acceptable runtimes, they fail to have good performance consistently.

The third group is which may be called "the most desirable" according to our criteria, that consists of the MCWR, OBCA, MEMB, and CBB algorithms. These algorithms are quite decent both in terms of speed and accuracy. Note that OBCA is the most accurate in this group but it has sometimes a relatively long running time, and in its current implementation it is not applicable if one is also interested in the created boxes, however, a valid node-box assignment could be done easily. In some sense, the greedy coloring algorithm also belongs to this third group but it is usually slightly slower than the other members of this group. From Figs. 6 and 7 we can observe that the performance of the MEMB and the greedy algorithms are very similar, which has been also shown by Wu et al. (2017).

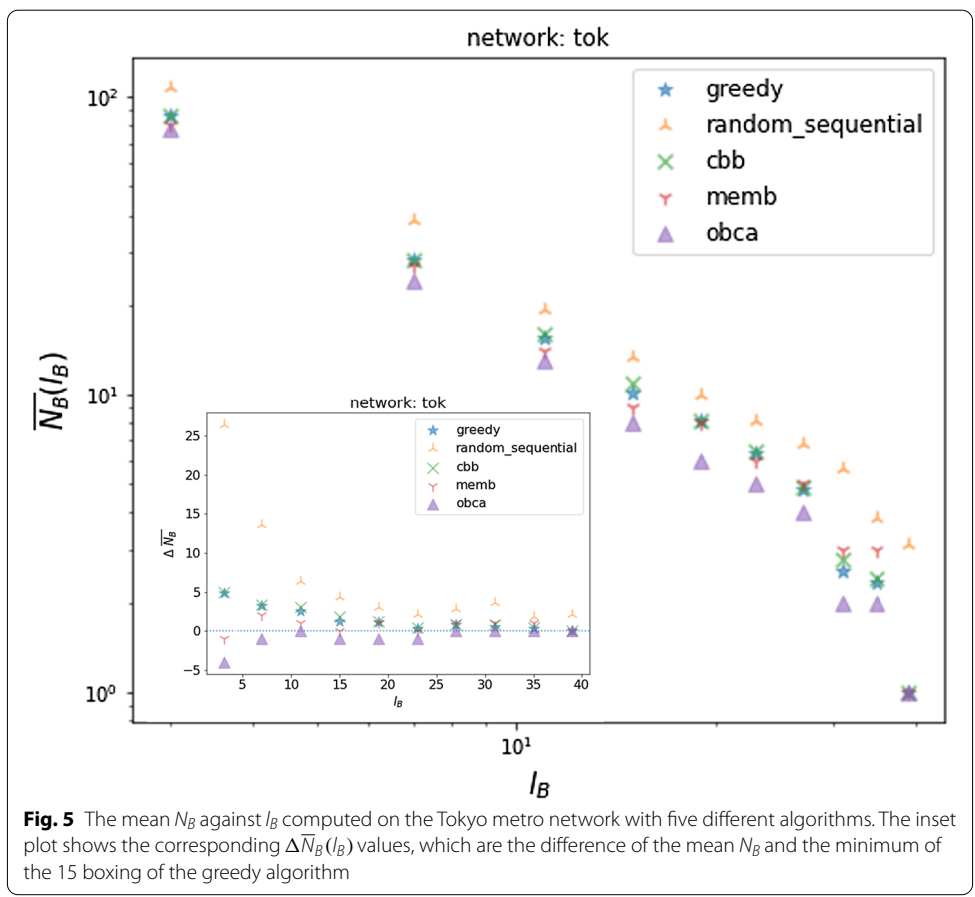

**Fig. 5** The mean $N_B$ against $l_B$ computed on the Tokyo metro network with five different algorithms. The inset plot shows the corresponding $\Delta \overline{N}_B(l_B)$ values, which are the difference of the mean $N_B$ and the minimum of the 15 boxing of the greedy algorithm

From Fig. 7 and Table 5, it is apparent that the random sequential and the REMCC algorithms have the highest variance in the performance scores, Moreover, the RS algorithm has also high intrinsic variance that has been also pointed out by Song et al. (2007).

One may be tempted to draw some general conclusions about the relation between the structure of the network and the performance of the algorithms, but it is a very challenging task considering the limited number of analyzed networks. However, without going into details about the qualitative behavior of the algorithms, we can notice that the "fast but inaccurate" random sequential and merge algorithms achieve one magnitude lower performance scores on the Minnesota road network than on the *C. elegans* network. We speculate that this may occur because the road network has concentrated degree distribution (most of the nodes have 2–4 neighbors), which is also shown by the low GINI

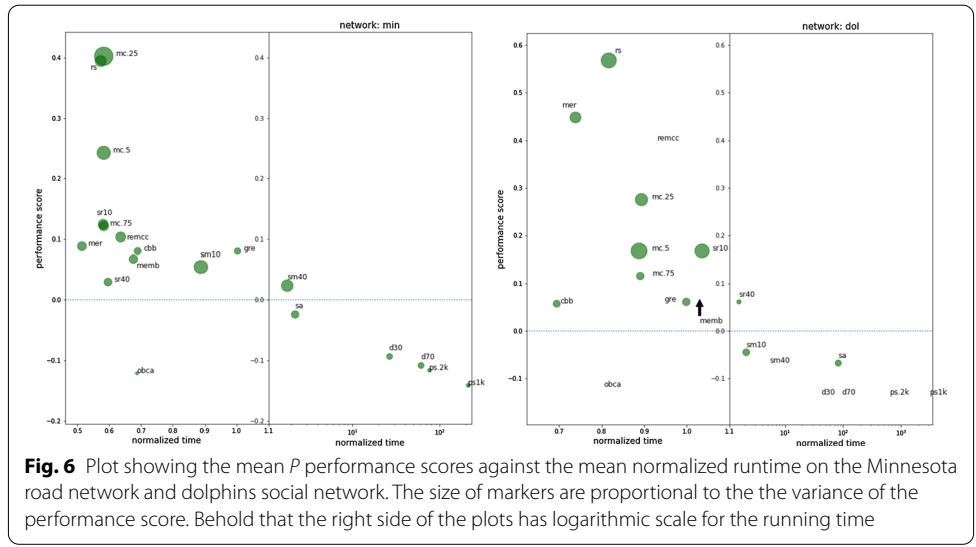

**Fig. 6** Plot showing the mean *P* performance scores against the mean normalized runtime on the Minnesota road network and dolphins social network. The size of markers are proportional to the the variance of the performance score. Behold that the right side of the plots has logarithmic scale for the running time

score (see Table 3), whereas in the *C. elegans* network, the degree distribution is much more uneven (higher GINI score).

**Comparing the estimated fractal dimensions**

As estimating the fractal dimension is one of the main points of box-covering algorithms, we also investigated what fractal dimensions the various algorithms yield on the investigated networks. Tables 6 and 8 suggest, that the dimension of the fractal networks is typically between 1.5 and 3.5. For example, the dimension of the Tokyo metro network is around 1.5, which makes sense, since the metro network is a composition of path graphs, however, at the transfer stations, these one-dimensional graphs are joined, which increases the fractal dimension of the network. On the other hand, clearly, a metro network does not contain as many junctions as the Minnesota road network, whose fractal dimension is around 2.0, as one would expect from a grid-like network.

We can observe that the estimated fractal dimension varies with the applied box-counting algorithm significantly. For example, on the *phd* network it ranges from 1.8 (REMCC) to 2.5 (merge) at the first experimenter and from 2.0 (REMCC) to 2.4 (merge) at the second experimenter. Similarly, the estimated fractal dimension of the Minnesota road network ranges from 1.6 (REMCC and SM) to 2.1 (merge) at the first experimenter and from 1.6 (SM) to 2.2 (DE and OBCA) at the second experimenter. Interestingly, the dimension estimated using the merge algorithm is generally much higher than the dimension resulted from the other algorithms. It is due to the fact that for small $l_B$ values the merge algorithm performs quite poorly but for larger $l_b$ values its performance increases, see Fig. 8.

In related works, the reported dimensions are close to our approximations. However, we found that the estimated fractal dimension not only depends on the applied algorithm but also on the way the regression is carried out, especially when the diameter of the network is small. For example, our first experimenter found that the $d_B$ dimension of the *dolphin* network is around 1.8, which is in alignment with the findings of Zheng et al. (2016) who found that the dimension is 1.86 using the MEMB algorithm. Deng et al. (2016) also reported similar values for *dolphin* network estimated by the MEMB, CBB, and greedy coloring algorithms: $d_B = 1.59$, $d_B = 1.90$, and $d_B = 1.83$, respectively. On the other hand, the estimated $d_B$ of the *dolphin* network according to our second experimenter is between 2.3 and 2.4, which is congruent with the result of Rosenberg (2020), who reported a dimension of 2.38. Hence, we can conclude that in the literature the reported fractal dimensions of the networks, especially of the small ones, are subject to the experimenters' subjective views, more specifically, what range of $l_b$ values is used for fitting.

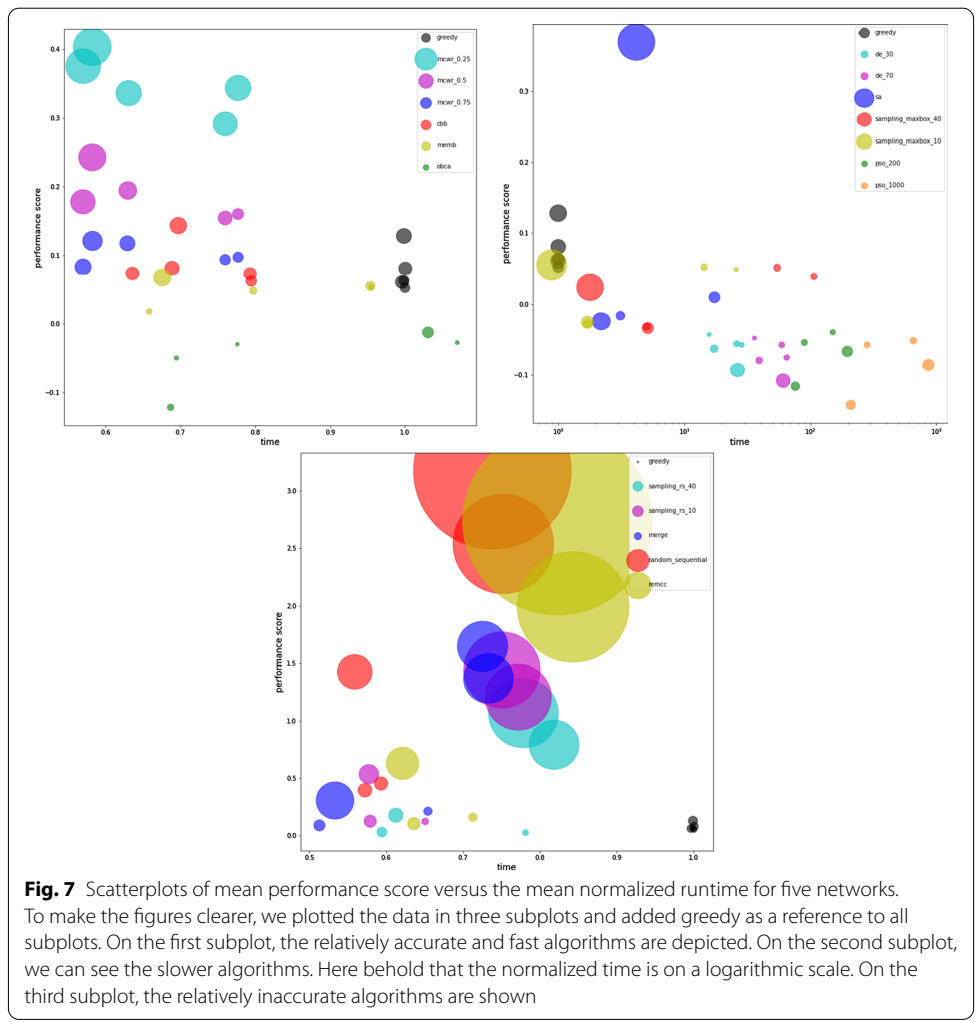

**Fig. 7** Scatterplots of mean performance score versus the mean normalized runtime for five networks. To make the figures clearer, we plotted the data in three subplots and added greedy as a reference to all subplots. On the first subplot, the relatively accurate and fast algorithms are depicted. On the second subplot, we can see the slower algorithms. Here behold that the normalized time is on a logarithmic scale. On the third subplot, the relatively inaccurate algorithms are shown

In the case of the other networks, the difference between the two experimenters is not so significant, however, the first experimenter's estimated values are usually below that of the second experimenter, which is probably due to the fact that the second experimenter fitted the line on a smaller range of $l_B$ values. This is also the reason why there are many missing data in the first experimenter's estimations.

According to our experimenters, most of the box-covering algorithms' results suggested that the fractal dimension of the *E. coli* network is around 3.5, which is close to

**Table 5** Variance in the *P* performance score: mean intrinsic/total standard deviation

|       | phd       | eco       | ful       | min       | tok       |
|-------|-----------|-----------|-----------|-----------|-----------|
| cbb   | 0.04/0.08 | 0.03/0.06 | 0.05/0.07 | 0.05/0.08 | 0.06/0.10 |
| d30   | 0.00/0.03 | 0.01/0.02 | 0.02/0.04 | 0.02/0.07 | 0.02/0.03 |
| d70   | 0.00/0.03 | 0.01/0.02 | 0.01/0.03 | 0.02/0.07 | 0.02/0.03 |
| gre   | 0.04/0.07 | 0.03/0.06 | 0.04/0.06 | 0.05/0.08 | 0.06/0.09 |
| mc.25 | 0.11/0.20 | 0.07/0.14 | 0.09/0.15 | 0.11/0.22 | 0.10/0.15 |
| mc.5  | 0.07/0.14 | 0.05/0.08 | 0.05/0.06 | 0.07/0.16 | 0.06/0.10 |
| mc.75 | 0.05/0.09 | 0.03/0.06 | 0.04/0.06 | 0.05/0.11 | 0.05/0.09 |
| memb  | 0.00/0.03 | 0.00/0.05 | 0.00/0.03 | 0.00/0.10 | 0.00/0.04 |
| mer   | 0.03/0.35 | 0.05/0.47 | 0.05/0.47 | 0.04/0.11 | 0.05/0.08 |
| obca  | 0.00/0.03 | 0.00/0.02 | 0.00/0.06 | 0.00/0.04 | 0.00/0.02 |
| ps.2k | 0.01/0.03 | –         | 0.01/0.03 | 0.02/0.04 | 0.01/0.05 |
| ps1k  | 0.00/0.03 | –         | 0.01/0.03 | 0.02/0.05 | 0.01/0.06 |
| remcc | 0.00/0.31 | 0.00/1.78 | 0.00/1.05 | 0.00/0.12 | 0.00/0.08 |
| rs    | 0.13/0.32 | 0.51/1.48 | 0.36/0.94 | 0.07/0.13 | 0.08/0.13 |
| sa    | 0.02/0.04 | –         | 0.06/0.19 | 0.05/0.09 | 0.04/0.06 |
| sm10  | 0.01/0.06 | 0.02/0.02 | 0.01/0.03 | 0.06/0.15 | 0.03/0.04 |
| sm40  | 0.01/0.06 | 0.01/0.03 | 0.01/0.03 | 0.06/0.14 | 0.02/0.04 |
| sr10  | 0.09/0.18 | 0.10/0.72 | 0.19/0.62 | 0.04/0.12 | 0.04/0.06 |
| sr40  | 0.08/0.14 | 0.23/0.65 | 0.20/0.47 | 0.05/0.09 | 0.04/0.06 |
|       | soc       | pol       | cel       | dol       |           |
| cbb   | 0.04/0.04 | 0.06/0.06 | 0.04/0.04 | 0.05/0.05 |           |
| d30   | 0.01/0.01 | 0.02/0.02 | 0.00/0.00 | 0.00/0.00 |           |
| d70   | 0.01/0.01 | 0.00/0.00 | 0.00/0.00 | 0.00/0.00 |           |
| gre   | 0.05/0.05 | 0.05/0.05 | 0.05/0.05 | 0.05/0.05 |           |
| mc.25 | 0.09/0.09 | 0.23/0.23 | 0.13/0.13 | 0.08/0.08 |           |
| mc.5  | 0.04/0.04 | 0.15/0.15 | 0.06/0.06 | 0.11/0.11 |           |
| mc.75 | 0.03/0.03 | 0.08/0.08 | 0.03/0.03 | 0.05/0.05 |           |
| memb  | 0.00/0.00 | 0.00/0.00 | 0.00/0.00 | 0.00/0.00 |           |
| mer   | 0.03/0.03 | 0.08/0.08 | 0.06/0.06 | 0.07/0.07 |           |
| obca  | 0.00/0.00 | 0.00/0.00 | 0.00/0.00 | 0.00/0.00 |           |
| ps.2k | 0.01/0.01 | 0.00/0.00 | 0.00/0.00 | 0.00/0.00 |           |
| ps1k  | 0.01/0.01 | 0.00/0.00 | 0.00/0.00 | 0.00/0.00 |           |
| remcc | 0.00/0.00 | 0.00/0.00 | 0.00/0.00 | 0.00/0.00 |           |
| rs    | 0.13/0.13 | 0.18/0.18 | 0.25/0.25 | 0.10/0.10 |           |
| sa    | 0.05/0.05 | 0.05/0.05 | 0.06/0.06 | 0.04/0.04 |           |
| sm10  | 0.03/0.03 | 0.06/0.06 | 0.00/0.00 | 0.05/0.05 |           |
| sm40  | 0.02/0.02 | 0.03/0.03 | 0.00/0.00 | 0.00/0.00 |           |
| sr10  | 0.10/0.10 | 0.12/0.12 | 0.45/0.45 | 0.10/0.10 |           |
| sr40  | 0.07/0.07 | 0.04/0.04 | 0.56/0.56 | 0.03/0.03 |           |

Values for mouse brain network (mou) are missing due to our criterion on the number of boxes. The hyphens indicate the lack of data due to the abort of the process the estimation of the related works: Zhang et al. (2014) used the greedy and the fuzzy algorithms and the estimated dimensions are $d_B = 3.54$ and $d_B = 3.54$ respectively. Similarly, Sun and Zhao (2014) reported $d_B = 3.47$ and $d_B = 3.35$ for the *E. coli* network based on the CBB and OBCA algorithms. Likewise, Locci et al. (2010) used the merge, the simulated annealing based, and the greedy coloring algorithms, and their

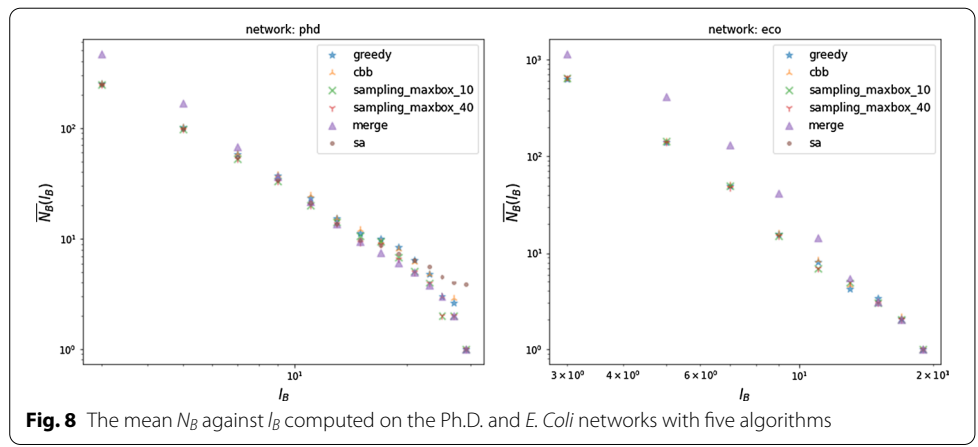

**Fig. 8** The mean $N_B$ against $l_B$ computed on the Ph.D. and *E. Coli* networks with five algorithms

**Table 6** Estimated $d_B$ values for the first experimenter

|  | phd | soc | mou | eco | pol | cel | ful | dol | min | tok |
|---|---|---|---|---|---|---|---|---|---|---|
| cbb | 1.9 | −1 | −1 | 3.4 | −1 | −1 | 3.3 | −1 | 1.8 | 1.3 |
| d30 | 2.2 | −1 | −1 | 3.5 | −1 | −1 | −1 | −1 | 1.9 | 1.4 |
| d70 | 2.1 | −1 | −1 | 3.4 | −1 | −1 | 3.4 | −1 | 1.9 | 1.4 |
| fuz | 2.3 | −1 | −1 | −1 | −1 | −1 | −1 | −1 | 1.9 | 1.9 |
| gre | 2.0 | −1 | −1 | 3.5 | −1 | −1 | 3.1 | −1 | 1.8 | 1.3 |
| mc.25 | 2.0 | −1 | −1 | 3.1 | −1 | 2.7 | 3.1 | 1.7 | 1.7 | 1.3 |
| mc.5 | 2.0 | −1 | −1 | 3.2 | −1 | −1 | 3.2 | 1.9 | 1.7 | 1.4 |
| mc.75 | 2.1 | −1 | −1 | 3.3 | −1 | −1 | 3.3 | −1 | 1.7 | 1.3 |
| memb | 2.0 | −1 | −1 | 3.5 | −1 | −1 | 3.2 | −1 | 1.7 | 1.3 |
| mer | 2.5 | −1 | −1 | 4.9 | −1 | −1 | −1 | −1 | 2.1 | 1.6 |
| obca | 2.1 | −1 | −1 | 3.4 | −1 | −1 | 3.2 | −1 | 1.8 | 1.4 |
| ps.2k | 2.1 | −1 | −1 | – | −1 | −1 | 3.6 | −1 | 1.8 | 1.3 |
| ps1k | 2.2 | −1 | −1 | – | −1 | −1 | 3.6 | −1 | 1.8 | 1.3 |
| remcc | 1.8 | −1 | −1 | −1 | −1 | −1 | 2.7 | 1.8 | 1.6 | 1.1 |
| rs | 2.2 | −1 | −1 | −1 | −1 | 3.4 | 3.0 | 2.1 | 1.8 | 1.3 |
| sa | 1.9 | −1 | −1 | – | −1 | −1 | 3.1 | −1 | 1.7 | 1.1 |
| sm10 | 2.0 | −1 | −1 | 3.5 | −1 | −1 | 3.6 | −1 | 1.6 | 1.4 |
| sm40 | 2.0 | −1 | −1 | 3.5 | −1 | −1 | 3.2 | −1 | 1.6 | 1.4 |
| sr10 | 2.1 | −1 | −1 | 3.4 | −1 | −1 | 4.0 | −1 | 1.8 | 1.4 |
| sr40 | 2.1 | −1 | −1 | 3.4 | −1 | −1 | 3.1 | −1 | 1.8 | 1.3 |

Here, −1.0 means that the network could not be considered fractal or there were too few datapoints. The hyphens indicate the lack of data due to the abort of the process estimations for the *E. coli* network are $d_B = 3.57$, $d_B = 3.47$, and $d_B = 3.44$ respectively. Finally, Deng et al. (2016) used the MEMB, CBB, and the greedy coloring algorithms and the corresponding estimated fractal dimensions for the *E. coli* network are $d_B = 2.95$, $d_B = 3.44$, and $d_B = 3.01$.

Sun and Zhao (2014) also estimated the fractal dimension of the *C. elegans* network with the OBCA and the CBB algorithms: $d_B = 2.99$ and $d_B = 2.95$, which are also close to the estimations of our experimenters, which is around 3.0.

**Table 7** SSE values for the fractal dimension estimation according to the first experimenter

|  | phd | soc | mou | eco | pol | cel | ful | dol | min | tok |
|---|---|---|---|---|---|---|---|---|---|---|
| cbb | 0.04 | − 1 | − 1 | 0.08 | − 1 | − 1 | 0.14 | − 1 | 0.06 | 0.02 |
| d30 | 0.08 | − 1 | − 1 | 0.09 | − 1 | − 1 | − 1 | − 1 | 0.09 | 0.03 |
| d70 | 0.07 | − 1 | − 1 | 0.09 | − 1 | − 1 | 0.16 | − 1 | 0.06 | 0.03 |
| fuz | 0.01 | − 1 | − 1 | − 1 | − 1 | − 1 | − 1 | − 1 | 0.04 | 0.13 |
| gre | 0.07 | − 1 | − 1 | 0.10 | − 1 | − 1 | 0.10 | − 1 | 0.06 | 0.02 |
| mc.25 | 0.04 | − 1 | − 1 | 0.07 | − 1 | 0.07 | 0.10 | 0.07 | 0.04 | 0.04 |
| mc.5 | 0.05 | − 1 | − 1 | 0.08 | − 1 | − 1 | 0.06 | 0.09 | 0.04 | 0.05 |
| mc.75 | 0.07 | − 1 | − 1 | 0.08 | − 1 | − 1 | 0.14 | − 1 | 0.04 | 0.05 |
| memb | 0.06 | − 1 | − 1 | 0.09 | − 1 | − 1 | 0.12 | − 1 | 0.07 | 0.04 |
| mer | 0.04 | − 1 | − 1 | 0.09 | − 1 | − 1 | − 1 | − 1 | 0.01 | 0.04 |
| obca | 0.06 | − 1 | − 1 | 0.09 | − 1 | − 1 | 0.14 | − 1 | 0.07 | 0.02 |
| ps.2k | 0.06 | − 1 | − 1 | – | − 1 | − 1 | 0.20 | − 1 | 0.07 | 0.03 |
| ps1k | 0.08 | − 1 | − 1 | – | − 1 | − 1 | 0.20 | − 1 | 0.05 | 0.03 |
| remcc | 0.03 | − 1 | − 1 | − 1 | − 1 | − 1 | 0.10 | 0.10 | 0.04 | 0.07 |
| rs | 0.05 | − 1 | − 1 | − 1 | − 1 | 0.15 | 0.19 | 0.08 | 0.03 | 0.02 |
| sa | 0.04 | − 1 | − 1 | – | − 1 | − 1 | 0.15 | − 1 | 0.03 | 0.04 |
| sm10 | 0.04 | − 1 | − 1 | 0.09 | − 1 | − 1 | 0.22 | − 1 | 0.03 | 0.03 |
| sm40 | 0.04 | − 1 | − 1 | 0.09 | − 1 | − 1 | 0.17 | − 1 | 0.02 | 0.03 |
| sr10 | 0.04 | − 1 | − 1 | 0.11 | − 1 | − 1 | 0.15 | − 1 | 0.05 | 0.07 |
| sr40 | 0.05 | − 1 | − 1 | 0.13 | − 1 | − 1 | 0.22 | − 1 | 0.05 | 0.03 |

Here, − 1.0 means that the network could not be considered fractal or there were too few datapoints. The hyphens indicate the lack of data due to the abort of the process

**Table 8** Estimated $d_B$ values according to the second experimenter

|  | phd | soc | mou | eco | pol | cel | ful | dol | min | tok |
|---|---|---|---|---|---|---|---|---|---|---|
| cbb | 2.0 | − 1 | − 1 | 3.5 | 2.4 | 3.0 | 3.5 | 2.4 | 1.9 | 1.5 |
| d30 | 2.2 | − 1 | − 1 | 3.5 | 2.1 | 3.0 | 3.6 | 2.3 | 2.2 | 1.5 |
| d70 | 2.2 | − 1 | − 1 | 3.5 | 2.1 | 3.0 | 3.6 | 2.3 | 2.2 | 1.5 |
| fuz | 2.3 | − 1 | − 1 | 2.9 | 1.4 | 1.5 | 2.9 | 1.4 | 1.9 | 2.0 |
| gre | 2.0 | − 1 | − 1 | 3.5 | 2.4 | 3.1 | 3.5 | 2.4 | 2.0 | 1.5 |
| mc.25 | 2.2 | − 1 | − 1 | 3.3 | 2.5 | 3.2 | 3.6 | 2.1 | 2.0 | 1.4 |
| mc.5 | 2.2 | − 1 | − 1 | 3.6 | 2.4 | 3.4 | 3.7 | 2.1 | 2.0 | 1.5 |
| mc.75 | 2.2 | − 1 | − 1 | 3.9 | 2.7 | 3.4 | 4.1 | 2.3 | 1.9 | 1.5 |
| memb | 2.3 | − 1 | − 1 | 3.9 | 2.7 | 3.6 | 4.2 | 2.4 | 2.0 | 1.4 |
| mer | 2.4 | − 1 | − 1 | 4.9 | 2.9 | 3.6 | 4.8 | 2.5 | 2.1 | 1.5 |
| obca | 2.1 | − 1 | − 1 | 3.5 | 2.3 | 3.0 | 3.9 | 2.4 | 2.2 | 1.5 |
| ps1k | 2.2 | − 1 | − 1 | – | 2.1 | 2.3 | 3.9 | 2.3 | 2.1 | 1.5 |
| ps.2k | 2.2 | − 1 | − 1 | – | 2.1 | 3.0 | 3.9 | 2.3 | 2.2 | 1.5 |
| remcc | 2.0 | − 1 | − 1 | 3.5 | 3.2 | 2.9 | 3.2 | 2.1 | 1.7 | 1.2 |
| rs | 2.4 | − 1 | − 1 | 4.1 | 2.9 | 4.2 | 3.7 | 2.2 | 2.0 | 1.5 |
| sa | 2.1 | − 1 | − 1 | – | 1.5 | 1.6 | 3.3 | 1.3 | 1.8 | 1.2 |
| sm10 | 2.1 | − 1 | − 1 | 3.6 | 2.4 | 3.0 | 4.0 | 2.4 | 1.6 | 1.5 |
| sm40 | 2.1 | − 1 | − 1 | 3.6 | 2.4 | 3.0 | 4.0 | 2.4 | 1.6 | 1.5 |
| sr10 | 2.4 | − 1 | − 1 | 4.3 | 2.6 | 4.4 | 4.4 | 2.3 | 2.1 | 1.4 |
| sr40 | 2.3 | − 1 | − 1 | 4.2 | 2.9 | 4.3 | 4.0 | 2.4 | 2.1 | 1.4 |

Here, − 1.0 means that the network could not be considered fractal or there were too few datapoints. The hyphens indicate the lack of data due to the abort of the process

**Table 9** SSE values for the fractal dimension estimation according to the second experimenter

| sse | phd | soc | mou | eco | pol | cel | ful | dol | min | tok |
|---|---|---|---|---|---|---|---|---|---|---|
| cbb | 0.05 | − 1 | − 1 | 0.10 | 0.31 | 0.22 | 0.23 | 0.29 | 0.05 | 0.11 |
| d30 | 0.08 | − 1 | − 1 | 0.15 | 0.50 | 0.22 | 0.18 | 0.07 | 0.04 | 0.12 |
| d70 | 0.08 | − 1 | − 1 | 0.12 | 0.50 | 0.22 | 0.17 | 0.07 | 0.05 | 0.12 |
| fuz | 0.01 | − 1 | − 1 | 0.31 | 0.22 | 0.34 | 0.28 | 0.18 | 0.04 | 0.10 |
| gre | 0.05 | − 1 | − 1 | 0.13 | 0.28 | 0.22 | 0.24 | 0.32 | 0.10 | 0.13 |
| mc.25 | 0.04 | − 1 | − 1 | 0.13 | 0.25 | 0.04 | 0.09 | 0.13 | 0.04 | 0.07 |
| mc.5 | 0.04 | − 1 | − 1 | 0.15 | 0.32 | 0.20 | 0.07 | 0.15 | 0.05 | 0.08 |
| mc.75 | 0.04 | − 1 | − 1 | 0.08 | 0.60 | 0.11 | 0.19 | 0.10 | 0.05 | 0.08 |
| memb | 0.05 | − 1 | − 1 | 0.13 | 0.70 | 0.24 | 0.23 | 0.16 | 0.07 | 0.08 |
| mer | 0.07 | − 1 | − 1 | 0.10 | 0.11 | 0.20 | 0.22 | 0.11 | 0.01 | 0.02 |
| obca | 0.07 | − 1 | − 1 | 0.13 | 0.26 | 0.22 | 0.27 | 0.07 | 0.03 | 0.12 |
| ps1k | 0.08 | − 1 | − 1 | – | 0.50 | 0.22 | 0.25 | 0.07 | 0.05 | 0.13 |
| ps.2k | 0.07 | − 1 | − 1 | – | 0.50 | 0.22 | 0.25 | 0.07 | 0.04 | 0.13 |
| remcc | 0.08 | − 1 | − 1 | 0.19 | 0.36 | 0.27 | 0.13 | 0.15 | 0.05 | 0.05 |
| rs | 0.04 | − 1 | − 1 | 0.12 | 0.30 | 0.23 | 0.19 | 0.12 | 0.02 | 0.07 |
| sa | 0.09 | − 1 | − 1 | – | 0.11 | 0.26 | 0.15 | 0.11 | 0.04 | 0.06 |
| sm10 | 0.05 | − 1 | − 1 | 0.12 | 0.58 | 0.22 | 0.27 | 0.07 | 0.03 | 0.12 |
| sm40 | 0.05 | − 1 | − 1 | 0.12 | 0.58 | 0.22 | 0.27 | 0.07 | 0.02 | 0.12 |
| sr10 | 0.04 | − 1 | − 1 | 0.11 | 0.26 | 0.18 | 0.14 | 0.20 | 0.03 | 0.07 |
| sr40 | 0.04 | − 1 | − 1 | 0.14 | 0.41 | 0.20 | 0.18 | 0.28 | 0.04 | 0.05 |

Here, − 1.0 means that the network is not fractal or there are too few datapoints. The hyphens indicate the lack of data due to the abort of the process

While the first experimenter did not find the estimation eligible for the *Polbooks* network, according to our second experimenter, its box dimension is around 2.4, which is also in alignment with the findings of Deng et al. (2016), who reported the following fractal dimension values for the *Polbooks* network approximated by MEMB, CBB, and the greedy coloring algorithm: $d_B = 1.85$, $d_B = 2.29$, and $d_B = 2.27$, respectively.

Note that the Facebook social network and the brain network could not be considered fractal networks, since the $N_B$ decays rather exponentially with the size of the boxes, and the estimated dimension would be above 5 in these networks.

The related works do not report the error of the regression analysis, however, we believe that to gain a better understanding of the goodness-of-fit of the line, it is also important to measure the SSE values, which are shown in Tables 7 and 9. We can observe that these SSE values are relatively large that is due to the small number of observations, especially in the case of the second experimenter (Table 9). These results also support the fact that the exact choice of the range of the $l_B$ values where the linear regression is carried out, significantly influences both the value and the error of the estimated fractal dimension. We argue that the impact of the range on estimating the dimension might be greater than the impact of the applied box-covering algorithm itself, especially in the case of small networks. Unfortunately, there is no canonical method for determining the appropriate range where the fractal scaling holds, which calls for further research.

### Limitations and future work

In this review, we attempted to be as comprehensive as possible, but the dynamic improvements in the field of fractal networks and box-covering algorithms make it hardly possible to be fully comprehensive. However, since our implementation is open source, one can easily integrate novel algorithms into our package.

It is also fair to acknowledge that our analysis contains some arbitrary decisions. For example, the number of repetitions at each $l_B$ or the way we defined the acceptance region for box sizes might seem arbitrary. We believe, however, that for a meaningful comparison with a comprehensible performance measure, the aggregation of the results based on some arbitrary decisions is inevitable.

We only reported on the *total* variance of the *G* performance scores of the algorithm, but one can argue that in some situations the *intrinsic* variation is more informative to use. Furthermore, we assessed the runtime of the algorithms by averaging the normalized runtimes. In this setting, the relative runtimes have the same 'weight' for each box size what we believe is desirable but one could be more interested in the cases where the running time is quite long and not so much in the "easy and fast" cases.

Due to the shortcomings of the evaluation framework, this study alone is by no means authorized to give a final verdict on the algorithms. We believe that the proposed framework together with the open-source code base is an important contribution to the community and it can serve as a starting point for future investigations. To gain a better understanding and make a more confident judgment on the performance of the algorithms, a higher number of networks from various domains could be used, perhaps together with more illuminating performance measures.

Moreover, another interesting further direction of research is to study box-covering algorithms on mathematical network models.

### Abbreviations

RS: Random sequential; GC: Greedy coloring; MA: Merge algorithm; CBB: Compact-box burning; MBC: Modified box counting; MEMB: Max-excluded mass burning; REMCC: Ratio of excluded mass to closeness centrality; ECSA: Edge-covering with simulated annealing; SA: Simulated annealing; DE: Differential evolution; PSO: Particle swarm optimization; MOPSO: Multiple-objective particle swarm optimization; ACO: Ant colony optimization; OBCA: Overlapping box-covering algorithm; IOB: Improved overlapping box-covering; MVB: Minimal-value burning; SM: Sampling-based method; CL: Coulomb's Law; DBCA: Deterministic box-covering algorithm; COM: Community-structure-based; Abbr: Abbreviation; Ref: Reference; phd: CSPhd network; cel: *C. elegans* metabolic network; soc: Caltech36 social network from Facebook; ful: *A. Fulgidus* cellular network; pol: Polbooks network; dol: Dolphins social network; mou: Mouse brain network; min: Minnesota road network; eco: *E. coli* cellular network; tok: Tokyo metro network; SSE: Sum of squared errors.

### Acknowledgements
We are grateful to Botond Diviki-Nagy for taking part in the implementation of the algorithms.

### Authors' contributions
RM has conceived the study. RM and MN reviewed the literature and collected the algorithms. PTK and MN implemented the algorithms, PTK developed the framework of the algorithms, and PTK performed the analysis. PTK prepared the original draft. MN and RM reviewed and edited the manuscript. All authors have read and agreed to the published version of the manuscript.

### Authors' Information
Péter Tamás Kovács, is a Master student in mathematics at the Budapest University of Technology and Economics. He earned his B.Sc. degree in physics with highest honors.

Marcell Nagy, is Ph.D. student in applied mathematics at the Department of Stochastics, Budapest University of Technology and Economics. His research focuses on data-driven network analysis and educational data science. He is the deputy team leader of the Human and Social Data Science Lab.

Roland Molontay, is a research fellow at the MTA-BME Stochastics Research Group and an assistant professor of applied mathematics and data science at the Budapest University of Technology and Economics (BME). His research focuses on network theory and on data-driven educational research. He is the founder and leader of the Human and Social Data Science Lab at BME.

**Funding**
The research presented in this paper was supported by the Ministry of Innovation and the National Research, Development and Innovation Office within the framework of the Artificial Intelligence National Laboratory Programme. The work of M. Nagy is funded by the ÚNKP-20-3-I New National Excellence Program of the Ministry for Innovation and Technology, Hungary. The work of R. Molontay is supported by the NKFIH K123782 research grant and by the ÚNKP-21-4-II New National Excellence Program of the Ministry for Innovation and Technology, Hungary.

**Availability of data and materials**
The datasets analyzed in this study and the algorithms implemented in Python are publicly available at the following GitHub repository: https://github.com/PeterTKovacs/boxes.

## Declarations

**Competing interests**
The authors declare that they have no competing interests.

**Author details**
[1]Department of Stochastics, Budapest University of Technology and Economics, Budapest, Hungary. [2]MTA-BME Stochastics Reseach Group, Budapest, Hungary.

**Publisher's Note**